\documentclass[aps,noshowpacs,groupedaddress,floatfix,nofootinbib,12pt]{revtex4}

\usepackage{amsmath}
\usepackage{amsfonts}
\usepackage{amssymb}
\usepackage{array}
\usepackage{braket}

\usepackage{color,hyperref}
\usepackage{graphicx}
\usepackage{liealg}
\usepackage{longtable}

\usepackage[flushleft]{threeparttable}  

\usepackage{mcmath}
\newcommand{\tme}[3]{\braket{#1|#2|#3}}

\newcommand{\calL}{\mathcal{L}}
\newcommand{\calW}{\mathcal{W}}

\newcommand{\Ad}{A^\dagger}    
\newcommand{\Ax}{A}

\newcommand{\Adn}{\mathcal{A}^\dagger}
\newcommand{\Axn}{\mathcal{A}}
\newcommand{\Adnt}{\tilde{\mathcal{A}}^\dagger}
\newcommand{\Axnt}{\tilde{\mathcal{A}}}

\begin{document}
\title{Algebraic evaluation of matrix elements in the Laguerre function basis}
\author{A. E. McCoy}
\affiliation{Department of Physics, University of Notre Dame,
Notre Dame, Indiana 46556-5670, USA}
\author{M. A. Caprio}
\affiliation{Department of Physics, University of Notre Dame,
Notre Dame, Indiana 46556-5670, USA}

\date{\today}
\begin{abstract}
The Laguerre functions constitute one of the fundamental basis sets
for calculations in atomic and molecular electron-structure theory,
with applications in hadronic and nuclear theory as well.  While
similar in form to the Coulomb bound-state eigenfunctions (from the
Schr\"odinger eigenproblem) or the Coulomb-Sturmian functions (from a
related Sturm-Liouville problem), the Laguerre functions, unlike these
former functions, constitute a complete, discrete, orthonormal set for
square-integrable functions in three dimensions. We construct the
$\grpsu{1,1}\times\grpso{3}$ dynamical algebra for the Laguerre
functions and apply the ideas of factorization (or supersymmetric
quantum mechanics) to derive shift operators for these functions.  We
use the resulting algebraic framework to derive analytic expressions
for matrix elements of several basic radial operators (involving
powers of the radial coordinate and radial derivative) in the Laguerre
function basis.  We illustrate how matrix elements for more general
spherical tensor operators in three dimensional space, such as the
gradient, may then be constructed from these radial matrix elements.
\end{abstract}

\sloppy



\maketitle

\section{Introduction}
\label{sec-intro}

The Laguerre
functions~\cite{jcp-23-1955-1362-Shull,jmp-21-1980-2725-Filter,jmp-26-1985-276-Weniger}
constitute one of the fundamental basis sets for calculations in
atomic and molecular electron-structure
theory~\cite{jcp-23-1955-1362-Shull,jcp-23-1955-1565-Shull,pr-101-1956-1730-Lowdin,helgaker2000:electron-structure},
with application in
hadronic~\cite{prd-33-1986-3338-Jacobs,prd-47-1993-4122-Fulcher,prd-49-1994-4675-Olson,prd-51-1995-5079-Olsson,
  jcp-134-1997-231-Keister,pervin2005:diss} and nuclear
structure~\cite{prc-84-2011-034003-Veerasamy,prc-86-2012-034312-Caprio,prc-90-2014-034305-Caprio}
theory as well.  The Laguerre functions share the same functional form
as the Coulomb bound-state wave functions (obtained as solutions to
the Schr\"odinger equation central force eigenproblem) or the
Coulomb-Sturmian functions (obtained as solutions to a related
Sturm-Liouville problem).  However, unlike these functions, the
Laguerre functions constitute a complete, discrete, orthonormal set of
square-integrable functions in three dimensions.

Numerical solution of quantum mechanical problems~--- for the present
discussion, we can take this to mean one-body and many-body
Schr\"odinger equation problems, although the applicability is more
general~--- may be represented in terms of an expansion of the wave
function on a set of basis functions.  The Schr\"odinger equation
eigenproblem, when expanded in a discrete basis, is recast as a
Hamiltonian matrix eigenproblem~\cite{sakurai1994:qm}.  To proceed
with this approach, it is necessary to be able to evaluate the matrix
elements of the operators of the problem, taken with respect to the
expansion basis.  Relevant operators may include those contributing to
the Hamiltonian, transition operators, and operators representing
moments or other static observables.
  
For numerical solution of many-body problems, the many-body basis
functions are obtained as products of single-particle basis functions
(more precisely, either symmetrized or antisymmetrized products,
according to the statistics of the
problem~\cite{negele1988:many-particle}).  When the many-body problem
is rotationally invariant, it is natural to consider an expansion in
terms of single-particle basis functions which are themselves obtained
as solutions to a rotationally-invariant central force problem, for
instance, the three-dimensional harmonic oscillator
functions~\cite{moshinsky1996:oscillator}.  The solution sets to
central force problems are well known to factorize into radial and
angular functions, arising from the factorizability of the Hilbert
space of square-integrable functions on $\bbR^3$ as
$\calL^2(\bbR^3)=\calL^2(\bbR^+)\times\calL^2(S^2)$.  The angular
functions, on the sphere $S^2$, are the well-known spherical
harmonics~\cite{varshalovich1988:am}.  For each choice of angular
momentum $l$, the associated radial functions form a complete set of
square-integrable functions on $\bbR^+$.  Matrix elements of spherical
tensor operators (including rotationally-invariant scalar operators
and vector operators) may be simplified through a corresponding
factorization into radial and angular dependences, together with the
Wigner-Eckart theorem~\cite{varshalovich1988:am} for the angular
dependence.

Most basic among the central force problems are those for the harmonic
oscillator potential and the Coulomb (hydrogenic) potential.  Both
these problems have a rich group theoretical structure (reviewed in
Ref.~\cite{wybourne1974:groups}), which allows analytic expressions to
be obtained for matrix elements of many operators of interest, built
from the coordinate vector $\vec{r}$ and momentum vector
$\vec{p}=-i\hbar\del$.

Underlying the convenience of these bases is the presence of an
$\grpsu{1,1}\times\grpso{3}$ dynamical
group~\cite{wybourne1974:groups}, which reflects the factorization of
the problem into radial and angular coordinates~--- with $\grpsu{1,1}$
operating on the radial functions and $\grpso{3}$ on the angular
functions.  The $\grpsu{1,1}$ \textit{spectrum generating algebra}
relates radial wave functions with different radial quantum number
(node number) $n$, at the same angular momentum $l$.  This
$\grpsu{1,1}$ algebra has been widely studied for both the
oscillator~\cite{ijqc-12-1977-875-Cizek,ajp-70-2002-945-Cooke,jpa-38-2005-10181-Rowe}
and Coulomb~\cite{pra-3-1971-1546-Armstrong,
  ijqc-12-1977-875-Cizek,aqc-19-1987-1-Adams,jpa-26-1993-1601-Cooper,ajp-70-2002-945-Cooke,jpa-40-2007-7721-Kelbert}
problems.  The $\grpso{3}$ algebra, in turn, is the standard angular
momentum algebra, which relates angular wave functions (spherical
harmonics) with different angular momentum projection quantum number
$m$, at given $l$~\cite{varshalovich1988:am}.

While the $\grpsu{1,1}\times\grpso{3}$ algebra thus provides the
framework for relating states of the same angular momentum (and, in
fact, for calculating matrix elements among these states, at least for
certain suitable operators), an additional structure is required to
connect basis states of different angular momentum.  This connection
is provided either by the factorization method (first applied by
Schr\"odinger for the Coulomb
problem~\cite{pria-46-1940-9-Schrodinger} and subsequently generalized
by Infeld and Hull~\cite{rmp-23-1951-21-Infeld}) or, equivalently, through the ideas of
supersymmetric quantum mechanics (SUSY QM)~\cite{prep-251-1995-267-Cooper}.
\begin{figure}
\begin{center}
\includegraphics[width=0.8\hsize]{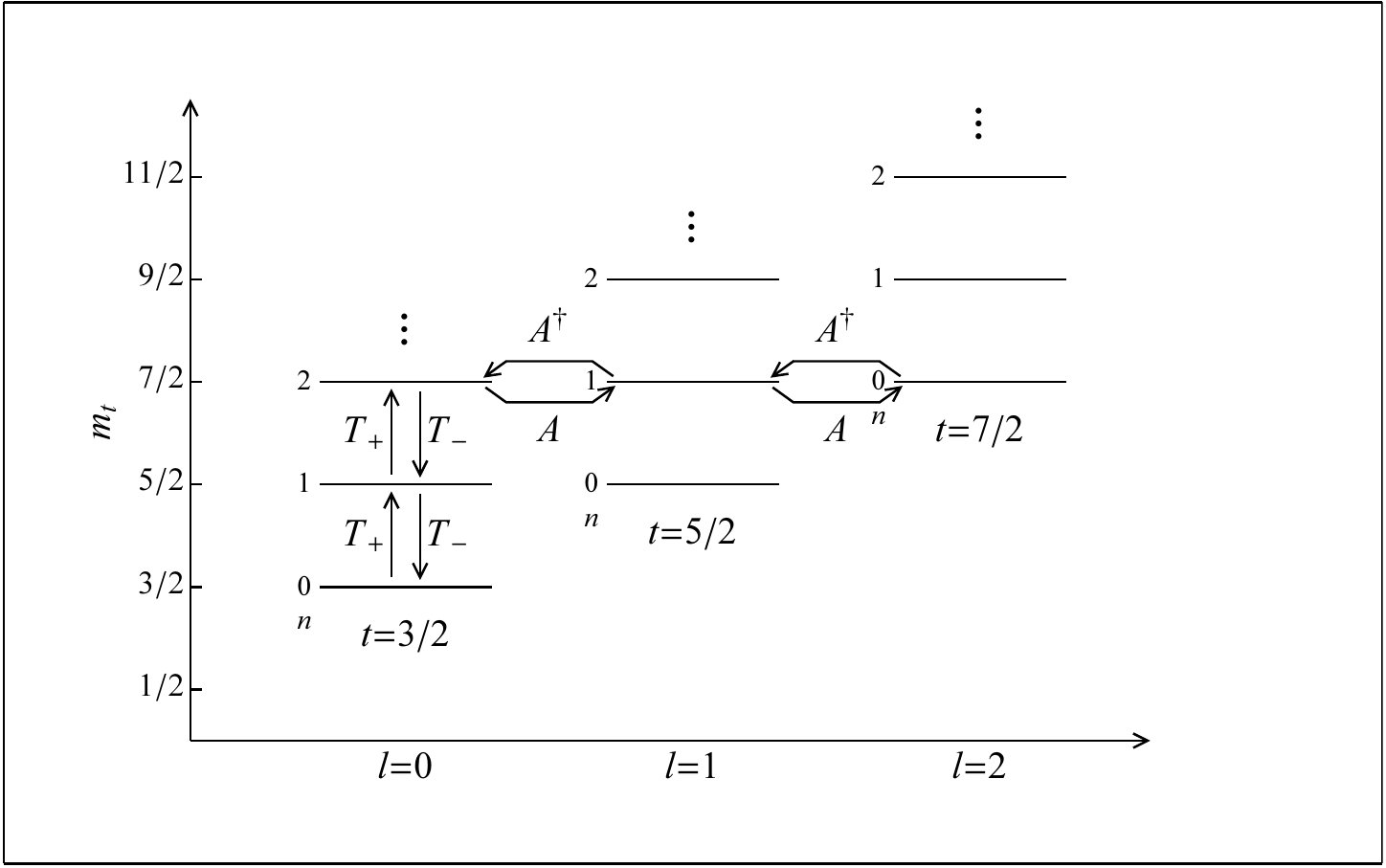}
\end{center}
\caption{Algebraic classification of basis functions, with respect to
  the $\grpsu{1,1}\times\grpso{3}$ dynamical algebra, and the
  pattern of laddering operators $T_\pm$ and shift operators, $A^\dagger$ and $A$, connecting
  these functions.  The quantum number $l$ labels the $\grpso{3}$
  irrep, and the dual quantum number $t$ labels the $\grpsu{1,1}$ irrep by its
lowest weight, where $m_t$ is the $\grpsu{1,1}$ weight (see text).}
\label{fig-lattice-su11}
\end{figure}
 
More concretely, the relations among basis states provided by the
algebraic structures are indicated in Fig.~\ref{fig-lattice-su11}.  To each
angular momentum $l$ is associated a set of radial wave functions.
These constitute a positive-discrete, infinite-dimensional irreducible
representation (irrep) of $\grpsu{1,1}$.  Radial wave functions
differing in radial quantum number $n$ by $\pm1$, at fixed $l$, are
related by the $\grpsu{1,1}$ \textit{ladder operators} $T_\pm$.  These
change the eigenvalue of the \textit{weight operator} $T_0$,
\textit{i.e.}, the weight $m_t=n+l+3/2$, by $\pm1$. Each $\grpsu{1,1}$
irrep is labeled by its lowest weight $t=l+3/2$.  Thus, to each irrep
(with label $l$) of the angular algebra $\grpso{3}$ there is uniquely
associated a dual irrep (with label $t=l+3/2$) of the $\grpsu{1,1}$
radial algebra~\cite{rmp-84-2012-711-Rowe}.  Then, \textit{shift
  operators}, $A^\dagger$ and $A$~\cite{jpa-26-1993-1601-Cooper}, obtained by the factorization or SUSY QM
approaches alluded to above, relate radial wave functions
corresponding to different angular momenta.  Specifically, the shift
operators connect radial wave functions of the same $m_t$ but
differing in $t$ by $\pm1$, \textit{i.e.}, radial wave functions
associated with angular functions of angular momentum $l$ differing by
$\pm1$.  The algebraic structure (ladder operators) and shift
operators in combination are what enable the evaluation of matrix
elements of operators in the full three-dimensional
space~\cite{wybourne1974:groups}.

 Let us now reexamine the choice of basis functions.  Since we shall
restrict attention to functions with radial-angular factorization,
these would also commonly be termed \textit{orbitals}.  In actual
calculations, the basis must be truncated to finite size (either the
set of single-particle basis functions may be truncated directly, or
some more general limitation may be placed on the set of many-body antisymmetrized
product states constructed from this basis~\cite{ppnp-69-2013-131-Barrett}).
Therefore, we must be concerned not only with mathematical
completeness of the basis but also with how well matched the basis is
to the physical problem at hand.  That is, the basis functions should
be chosen according to the following criteria (see
Ref.~\cite{helgaker2000:electron-structure}): (i)~The functions should
allow for a systematic approach to completeness as the single-particle
basis set is expanded.  (ii)~The functions should provide for rapid
convergence of the calculated energies or eigenfunctions to the true
results, requiring only a small number of basis functions for an
accurate description.  (iii)~It is also practically desirable that the
functions have a simple analytic form which permits convenient evaluation
of relevant operator matrix elements.  (Ideally, the functions should
also be orthogonal and easily normalized, to simplify the formulation
of the many-body problem, but this is not essential.)

The harmonic oscillator eigenfunctions are of the form (for
simplicity, we omit normalization factors and set the radial length parameter
to unity)~\cite{moshinsky1996:oscillator,jmp-26-1985-276-Weniger}
\begin{equation}
\label{eqn-ho-radial}
\Omega_{nlm}(\vec{r})\,\propto\,
r^{l+1}\,
L_n^{l+1/2}(r^2)\,
e^{-r^2/2}\,
Y_{lm}(\uvec{r}),
\end{equation}
where $L_n^\alpha(x)$ is the associated Laguerre polynomial of degree
$n$ and order $\alpha$, \textit{i.e.}, for parameter $\alpha$ in the associated Laguerre
equation~\cite{arfken2005:mathmethods}, and $Y_{lm}(\uvec{r})$ is a
spherical harmonic.  These functions constitute a complete,
orthonormal set on $\calL^2(\bbR^3)$.  However, while the
Schr\"odinger equation solutions for a finite-range potential
(or for an infinite-range
potential of Coulomb $1/r$ type) exhibit an exponential asymptotic
decay at large $r$ ($\propto e^{-\beta r}$), the oscillator functions
have Gaussian asymptotics ($\propto e^{-\alpha r^2}$), as an artifact of the
infinite (quadratic) binding potential of the harmonic oscillator
problem.  The oscillator functions are thus poorly matched to the
physical solutions in the asymptotic region and may therefore violate criterion~(ii) above for a suitable
basis, \textit{i.e.}, rapid convergence. In nuclear physics, for instance, while structure
involving only deeply-bound orbitals might be adequately represented
using basis functions with Gaussian asymptotics, weakly-bound orbitals
with distinctly exponential asymptotics influence many aspects of
nuclear structure, especially in light or neutron-rich nuclei (see,
\textit{e.g.}, Fig.~4 of Ref.~\cite{davies1966:hartree-fock} and
Fig.~4 of Ref.~\cite{jonson2004:light-dripline}).  

 The Coulomb wave functions, as solutions to a problem with finite
 binding, display the appropriate exponential asymptotics.  The
 bound-state (or negative-energy) Coulomb functions~\cite{jmp-26-1985-276-Weniger,arfken2005:mathmethods,helgaker2000:electron-structure} are of the
 form~\cite{fn-qn-notation}
\begin{equation}
\label{eqn-coulomb-radial}
W_{nlm}(\vec{r})\,\propto\,
\biggl(\frac{2r}{n+l+1}\biggr)^l\,
L_n^{2l+1}\biggl(\frac{2r}{n+l+1}\biggr)\,
e^{-r/(n+l+1)}\,
Y_{lm}(\uvec{r}).
\end{equation}
The \textit{full} set of solutions to the Schr\"odinger equation with
the Coulomb potential, including both these bound-state solutions and
the continuum (or positive-energy) solutions, constitutes a complete
set on $\calL^2(\bbR^3)$.  However, the bound-state solutions by
themselves do not.  This is not merely a formal deficiency: attempting
to use them as an expansion basis is well-known to result in
convergence to erroneous energies and wave
functions~\cite{jcp-23-1955-1362-Shull}.  It is therefore necessary to
supplement the bound-state solutions with the continuum functions.
The inconvenience of including continuum functions limits the
practical applicability of the Coulomb basis: both in terms of
arranging a systematic approach to completeness with increasing basis
size [criterion~(i)] and due to the complicated nature of the
continuum wave functions
[criterion~(iii)]~\cite{jmp-26-1985-276-Weniger,helgaker2000:electron-structure}.

One may instead seek a \textit{discrete} complete set, retaining the
same exponential asymptotics as the Coulomb wave functions, by
modifying the Schr\"odinger equation into a Sturm-Liouville problem
associated with the Coulomb potential (as detailed in
Sec.~\ref{Laguerre functions}), thereby obtaining the Coulomb-Sturmian
functions~\cite{zp-48-1928-469-Hylleraas,apny-19-1962-262-Rotenberg,tchimacta-44-1977-27-Klahn,jmp-26-1985-276-Weniger,fn-sturmian}.
These have been applied as expansion bases in atomic and molecular
theory~\cite{zp-48-1928-469-Hylleraas,apny-19-1962-262-Rotenberg,aamp-6-1970-233-Rotenberg,pra-9-1974-1209-Heller,aqc-67-2013-73-Coletti},
nuclear
scattering~\cite{prc-38-1988-2457-Papp,mplb-22-2008-2201-Papp}, and
other few-body problems~\cite{jpa-20-1987-153-Papp}.  The
Coulomb-Sturmian functions are of the form
\begin{equation}
\label{eqn-cs-radial}
\Psi_{nlm}(\vec{r})\,\propto\,
(2r)^l\,
L_n^{2l+1}(2r)\,
e^{-r}\,
Y_{lm}(\uvec{r}),
\end{equation}
that is, of the same functional form as the Coulomb eigenproblem
solutions~(\ref{eqn-coulomb-radial}) but independently rescaled in the
radial coordinate so that $r/(n+l+1)\rightarrow r$.  These radial
functions constitute a discrete set, complete on the Sobolev space
$\calW_2^{(1)}(\bbR^3)$ [\textit{i.e.}, the space of all functions in
  $\calL^2(\bbR^3)$, the generalized second derivatives of which are
  also in $\calL^2(\bbR^3)$]~\cite{jmp-26-1985-276-Weniger}, which
implies completeness on
$\calL^2(\bbR^3)$~\cite{tchimacta-44-1977-9-Klahn}.  However, these
functions are orthogonal with respect to the Coulomb potential $1/r$
as weighting function, rather than the usual Euclidean metric.
  
We therefore finally turn to the Laguerre functions~\cite{fn-lgf-radial,fn-lgf-associated}  which are of the form
\begin{equation}
\label{eqn-lg-radial}
\Lambda_{nlm}(\vec{r})\,\propto\,
(2r)^l\,
L_n^{2l+2}(2r)\,
e^{-r}\,
Y_{lm}(\uvec{r}).
\end{equation}
They are identical in form to the Coulomb-Sturmian
functions~(\ref{eqn-cs-radial}), differing only in the parameter
$\alpha$ of the associated Laguerre polynomial: odd $\alpha=2l+1$ for
the Coulomb-Sturmian functions \textit{vs.}~even $\alpha=2l+2$ for the
Laguerre functions~\cite{fn-lgf-vs-cs}.  The Laguerre functions satisfy
all the posited criteria: consitituting a discrete set, complete on
$\calL^2(\bbR^3)$, and orthogonal with respect to the usual Euclidean
metric, but with the physically-appropriate exponential asymptotics.
  
We demonstrate that an algebraic framework, analogous to that obtained
for the usual central force eigenfunctions, can likewise be
constructed for the Laguerre functions.  These results permit analytic
expressions to be obtained for matrix elements.  The Laguerre
functions are first introduced (Sec.~\ref{Laguerre functions}).  After
a brief review of the $\grpso{3}$ and $\grpsu{1,1}$ Lie algebras and
the structure of their irreps (Sec.~\ref{alg review}), we develop the
$\grpsu{1,1}$ algebra of the Laguerre radial functions
(Sec.~\ref{alg}), and we use this algebra to obtain analytic
expressions for the action of several basic operators (involving
powers of the radial coordinate $r$ and radial derivative operator
$d/dr$) on the basis of Laguerre radial functions~--- or,
equivalently, analytic expressions for matrix elements of these
operators (Sec.~\ref{radial MEs}).  The results thus obtained apply to
functions within a single $\grpsu{1,1}$ irrep, \textit{i.e.}, to the
set of radial functions associated with the same angular momentum.  We
then extend the ideas of factorization (or SUSY QM) to a quantum
number dependent formulation (Sec.~\ref{Ndep}), to construct shift
operators relating Laguerre radial functions in different irreps
(Sec.~\ref{CSshift}), and thus to obtain analytic expressions for the
matrix elements of radial operators between different irreps
(Sec.~\ref{CSshiftme}).  Finally, returning to the full Laguerre
functions in three dimensions, we demonstrate how the results for
radial matrix elements and shift operators are combined to yield
reduced matrix elements of spherical tensor operators, taking the
Laplacian and gradient operators for illustration (Sec.~\ref{lambda}).
  
\section{Laguerre functions\label{Laguerre functions}}

The Coulomb functions~\eqref{eqn-coulomb-radial}  are the solutions to the Schr\"odinger equation
\begin{equation}
\left(-\nabla^2-Z\frac{2}{r}\right)W({\bf r})=2E_{nl}W({\bf r}),\qquad
E_{nl}=-\frac{Z^2}{2(n+l+1)^2}.  \label{eqn-coulomb-de}
\end{equation}
As outlined above, these do \emph{not} form a complete set on $\mathcal{L}^2(\mathbb{R}^3)$
without including the continuum \cite{jcp-23-1955-1362-Shull,aamp-6-1970-233-Rotenberg}, making them an impractical choice
of basis.  

There is, however, a set of functions, closely related to
the Coulomb functions, called the Coulomb-Sturmian functions
\eqref{eqn-cs-radial}, which combine the discrete character of the
harmonic oscillator functions with the desired large $r$ behavior.
These functions are related to the Coulomb functions by making the
substitution
\begin{equation}
\frac{n+l+1}{Z}\rightarrow b,
\end{equation}
which recasts the differential equation~\eqref{eqn-coulomb-de} into
the form of a Sturm-Liouville equation
\begin{equation}
\left(-\nabla^2-\frac{2}{b}\frac{\beta_{nl}}{r}\right)\Psi({\bf r})=-\frac{1}{b^2}\Psi({\bf r}),\qquad \beta_{nl}=n+l+1, \label{Phide}
\end{equation}
and yields the Coulomb-Sturmian solutions
\begin{equation}
\Psi_{nlm}({\bf r})=\left(\frac{(2/b)^3n!}{2(n+l+1)(n+2l+1)!}\right)^{\frac12}\left(\frac{2r}{b}\right)^lL_n^{2l+1}(2r/b)e^{-r/b}Y_{lm}(\hat{r}).
\end{equation}
Observe that, while in the Schr\"odinger equation
eigenproblem~\eqref{eqn-coulomb-de} one varies the energy $E_{nl}$ to
satisfy the boundary conditions, in the Sturm-Liouville
eigenproblem~\eqref{Phide} one now holds the energy constant and
varies the coefficient $\beta_{nl}$, which governs the strength or
depth of the potential, to satisfy the boundary
conditions~\cite{apny-19-1962-262-Rotenberg}.
From the basic theory of the Sturm-Liouville equation~\cite{fn-sturm-liouville}, it follows that the
solutions $\Psi_{nlm}$ are orthogonal with respect to the Coulomb
potential $1/r$ as weighting
function~\cite{apny-19-1962-262-Rotenberg}.  These functions
$\Psi_{nlm}$ are also complete in the Sobolev space
$\calW_2^{(1)}(\mathbb{R}^3)$~\cite{jmp-26-1985-276-Weniger}.

However,
for calculations in physics it is useful to work with functions that
are orthonormal with respect to the usual integration metric $d^3r$
and complete on $\mathcal{L}^2(\mathbb{R}^3)$.  To obtain functions
with this property, the integration weight and norm are absorbed
into the function by multiplying $\Psi_{nlm}$ by
$[b(n+l+1)/r]^{1/2}$, and the shift $l\rightarrow l+1/2$ is
introduced~\cite{jmp-26-1985-276-Weniger}.
The resulting functions are the Laguerre functions,
\begin{equation}
\Lambda_{nlm}({\bf r})=\left(\frac{2}{b}\right)^{\frac32}\left(\frac{n!}{(n+2l+2)!}\right)^{\frac12}\left(\frac{2r}{b}\right)^lL_n^{2l+2}(2r/b)e^{-r/b}Y_{lm}(\hat{r})\label{Lambda},
\end{equation}
which satisfy the radial differential equation
\begin{equation}
\left[-\frac{\partial^2}{\partial
    r^2}-\frac{3}{r}\frac{\partial}{\partial
    r}+\frac{l(l+2)}{r^2}-\frac{2}{b}\frac{\alpha_{nl}}{r}\right]\Lambda_{nlm}({\bf
  r})=-\frac{1}{b^2}\Lambda_{nlm}({\bf
  r}),
\qquad\alpha_{nl}=n+l+\tfrac32.
\label{SCSde}
\end{equation}
The Laguerre functions
$\Lambda_{nlm}({\bf r})$ form a complete, discrete basis for
$L^2(\mathbb{R}^3 )$~\cite{tchimacta-44-1977-9-Klahn}, have the correct asymptotic behavior and are
orthogonal with respect to the $\mathbb{R}^3$ metric.  That is,
\begin{equation}
\int d^3 r\Lambda_{n'l'm'}({\bf r})\Lambda_{nlm}({\bf r})=\delta_{n,n'}\delta_{l,l'}\delta_{m,m'}.
\end{equation}
It is now the matrix elements calculated in this Laguerre basis that are of interest.  Certain matrix elements can be calculated via direct integration~\cite{prd-47-1993-4122-Fulcher, prd-51-1995-5079-Olsson}.  However, in this paper we use the factorability of the Hilbert space to develop a simpler and more elegant algebraic method for calculating matrix elements. 

The Laguerre function $\Lambda_{nlm}({\bf r})$ can be factorized into radial and angular functions as 
\begin{equation}
\Lambda_{nlm}({\bf r})=r^{-1}S_{nl}(r)Y_{lm}(\hat{\bf r})\label{sdef}.
\end{equation}
The integration weight has been absorbed into the radial function $S_{nl}(r)$, given by
\begin{equation}
S_{nl}(r)=\left(\frac{2}{b}\right)^{\frac12}\left(\frac{n!}{(n+2l+2)!}\right)^{\frac12}\left(\frac{2r}{b}\right)^{l+1}L_n^{2l+2}(2r/b)e^{-r/b}, 
\end{equation}
which satisfies the orthogonality condition
\begin{equation}
\int_0^\infty dr S_{n'l}(r)S_{nl}(r)=\delta_{n,n'}.
\end{equation}
The set of functions $\{S_{nl}|n=0,1,2,...\}$ forms a complete and orthogonal set  on $L^2(\mathbb{R}^+)$, for any $l=0,1,2,...$, independent of $b$ \cite{tchimacta-44-1977-27-Klahn}, so we make the simplifying substitution $b\rightarrow 1$ in the following discussion. The resulting functions, 
\begin{equation}
S_{nl}(r)=\left(\frac{2n!}{(n+2l+2)!}\right)^{\frac12}(2 r)^{l+1}L_n^{2l+2}(2 r)e^{- r},
\end{equation}
satisfy the differential equation 
\begin{equation}
\left[-\frac{d^2}{d r^2}-\frac1r\frac{d}{d r}+\frac{(l+1)^2}{r^2}-\frac{2\alpha_{nl}}{r}\right]S_{nl}(r)=-1S_{nl}(r),  \label{eqn-Snl-eigen}
\end{equation}
with $\alpha_{nl}$ as defined in~\eqref{SCSde}.
Although~\eqref{eqn-Snl-eigen} is not explicitly written in the form of a
Sturm-Liouville equation (see footnote~\cite{fn-sturm-liouville}), it nonetheless
may readily be put in this form by left multiplying by $r$.  For
convenience of notation, we will sometimes denote $\Lambda_{nlm}$ as
$\ket{nlm}$, $S_{nl}$ as $\ket{nl}$ and $Y_{lm}$ as $\ket{lm}$.

\subsection{Operator map to the radial basis\label{operator map}}
Although we want the matrix elements in the $\Lambda_{nlm}$ basis, the actual calculations will be carried out more simply using the functions  
\begin{equation}
\ket{nl}\ket{lm}\label{facth2},
\end{equation}
since they are orthogonal with respect to the measure  $dr\,d\Omega$ rather than the measure $r^2\,dr\,d\Omega$.  We start with the matrix element of a component of some irreducible, rank $a$ tensor operator
 \begin{equation}
 \mathcal{O}_{a\alpha}=R\Theta_{a\alpha},
 \end{equation}   
where $R$ is a scalar radial operator and $\Theta_{a}$ is a rank $a$ tensor with no radial dependence.  Although we are only considering operators that can be factored into radial and angular parts, many operators of interest can be rewritten as sums of such separable operators using various factorization lemmas, and so this method can also be used to calculate matrix elements of these operators.  As an example, matrix elements of the gradient operator ${\nabla_1}$ are calculated in Sec.~\ref{vec(nabla)}.  

The matrix element of the operator $\mathcal{O}_{a\alpha}$ separates into radial and angular factors as 
\begin{multline}
\braket{n'l'm'|\mathcal{O}_{a\alpha}|nlm}=\int d^3r\Lambda_{nlm}({\bf r})\mathcal{O}_{a\alpha}\Lambda_{nlm}({\bf r})\\
=\left[\int dr r^2r^{-1}S_{n'l'}(r)R(r)r^{-1}S_{nl}(r)\right]\left[\int d\Omega Y_{l'm'}(\hat{r})\Theta_{a\alpha}Y_{lm}(\hat{r})\right]
\end{multline}
That is, 
\begin{equation}
\braket{n'l'm'|\mathcal{O}_{a\alpha}|nlm}=\braket{n'l'|\gamma(R)|nl}\braket{l'm'|\Theta_{a\alpha}|lm},\label{matrixrel}
\end{equation}
where 
\begin{equation}
\gamma(R)=rR r^{-1}\label{symtrans}.
\end{equation}  
If the operator is purely radial ($\Theta_{a\alpha}\rightarrow{\bf 1}_{00}$), then we have that
\begin{equation}
\braket{n'l'm'|R|nlm}=\braket{n'l|\gamma(R)|nl}.  
\end{equation}
The similarity transformation given by \eqref{symtrans} will allow us to
calculate matrix elements in the $\Lambda_{nlm}$ basis by factoring
the matrix element into linear combinations of products of a matrix
element in the $S_{nl}$ basis and a matrix element in the well known
angular momentum basis,  as shown (for the case of positive weights) in Sec.~\ref{lambda}.

\section{Algebraic structure of Laguerre basis\label{algstruct}}
The factorization \eqref{sdef} of the $\Lambda_{nlm}$ basis for $\mathcal{L}(\mathbb{R}^3)$ has the underlying symmetry SU(1,1)$\times$SO(3), where SU(1,1) is the symmetry of the radial functions and SO(3) in the symmetry of $Y_{lm}$ angular functions.  It is well known that the SO(3) symmetry can be used to calculate matrix elements of operators on the $Y_{lm}$ basis.  In this section we discuss how SU(1,1) symmetry can be used to calculate the matrix elements on the $S_{nl}$ basis.   
\subsection{A brief review of the SO(3) and SU(1,1) Lie algebras\label{alg review}}
The Lie algebra SU(1,1) has a very similar structure to that of the Lie algebra SO(3), so we begin by summarizing the properties of the more familiar SO(3) algebra for comparison with the SU(1,1) algebra.  

The group SO(3) is the compact Lie group of rotations in three dimensions.  Its algebra is generated  by the operators $L_3$ and $ L_{\pm}$, which  satisfy the commutation relations 
\begin{equation}
[L_3,L_\pm]=\pm L_{\pm}
\end{equation}
and 
\begin{equation}
[L_+,L_-]=2L_3,  \label{angularcom}
\end{equation}
where $L^\dagger_+=L_-$.  The eigenfunctions of $L_3$ are the spherical harmonics $\ket{lm}$, where 
\begin{equation}
L_3\ket{lm}=m\ket{lm}. 
\end{equation}
The laddering operators $L_\pm$ have the well known action on the spherical harmonics 
\begin{equation}
L_\pm \ket{lm}=\sqrt{(l\mp m)(l\pm m+1)}\ket{l,m\pm1}.
\end{equation}

\begin{figure}
\begin{center}
\includegraphics[width=0.8\hsize]{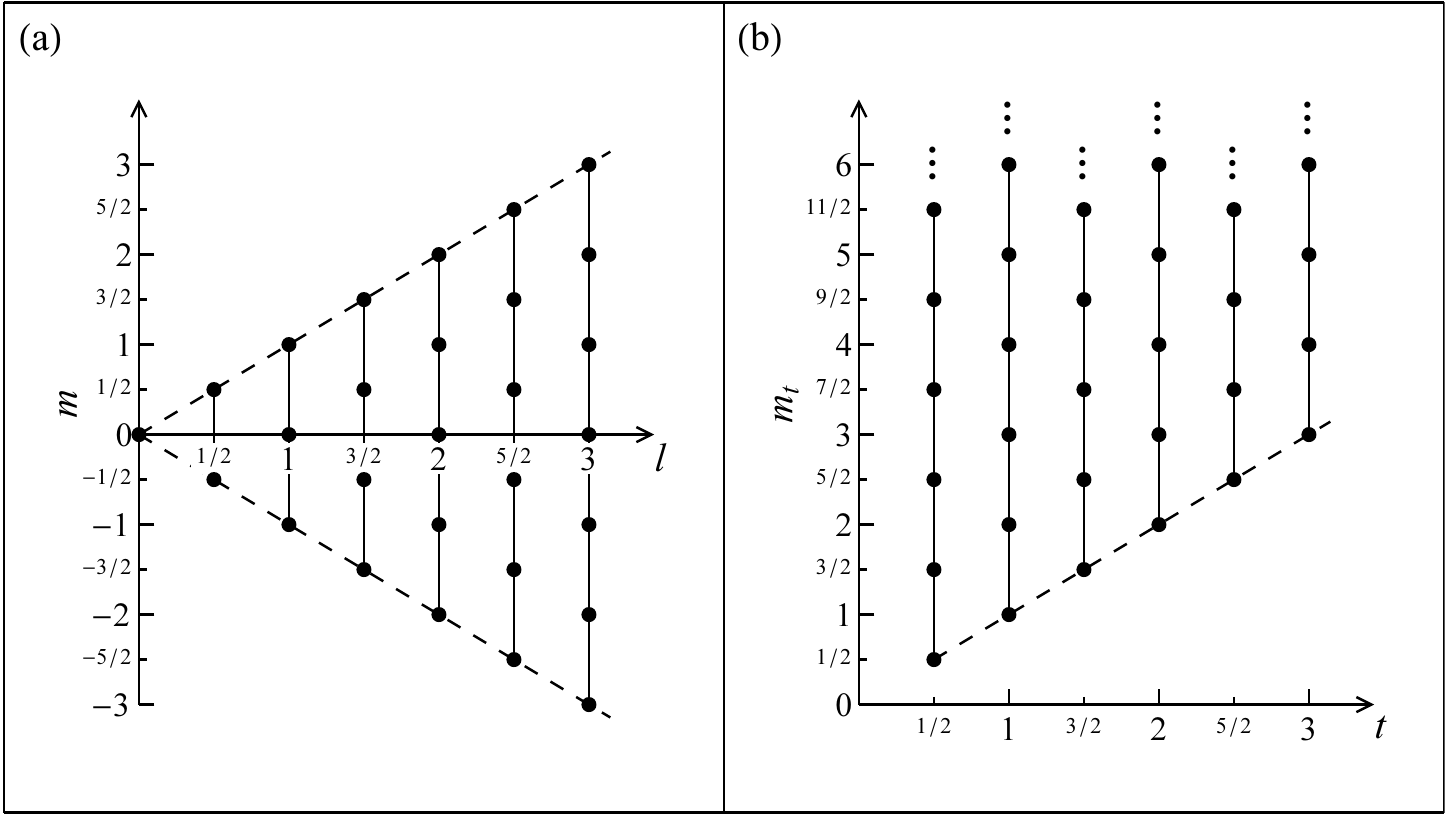}
\end{center}
\caption{Weights contained within (a)~$\grpso{3}$ and (b)~$\grpsu{1,1}$ irreps of the types considered in the present work (see text).  Weights ($m$ or $m_t$, respectively, for these two algebras) within an irrep
(labeled by extremal weight $l$ or $t$, respectively) are connected by solid lines.  The bounding weights, as functions of $l$ or $t$, are indicated by dashed lines.}
\label{fig-weights}
\end{figure}

Because SO(3) is a compact group, the weight $m$ is bounded above and below.  The bound is determined by the SO(3) Casimir operator  
\begin{equation}
C_{SO(3)}=L_3^2+\frac12\left(L_+L_-+L_-L_+\right),
\end{equation}
which acts on the eigenfunctions as 
\begin{equation}
C_{SO(3)}\ket{lm}=l(l+1)\ket{lm}. 
\end{equation}
The bounds above and below on $m$ are given by $\pm l$, that is  
\begin{equation}
m=l,l-1,...,-l,
\end{equation}
as shown in Fig.~\ref{fig-weights}(a). 

Similarly, the generators of SU(1,1) are $T_3$ and $T_\pm$, which  satisfy the commutation relations:
\begin{equation}
 [T_3,T_{\pm}]=\pm T_\pm\label{com1}
\end{equation}
and
\begin{equation}
[T_+,T_-]=-2T_3\label{negative},
\end{equation}
where $T_+^\dagger=T_-$.  
The actions of the generators on the eigenfunctions $\ket{tm_t}$ of $T_3$ are given by 
\begin{align}
T_3\ket{tm_t}&=m_t\ket{tm_t}\label{m}
\end{align}
and
\begin{align}
T_\pm\ket{tm_t}&=\sqrt{(t\pm m_t)(t\mp m_t-1)}\ket{t,m_t\pm1}\label{pm}.
\end{align}
The commutator in  \eqref{negative} differs from the commutator in
\eqref{angularcom} by a negative sign.  The group SU(1,1) is a non-compact group, and the weight $m_t$ is either bounded above or bounded below but takes on infinitely many values.   As with SO(3), the bound on $m_t$ is determined by the Casmir operator 
\begin{equation}
C_{SU(1,1)}=T_3^2-\frac12(T_+T_-+T_-T_+),
\end{equation}
 which acts on the eigenfunction as
\begin{equation}
C_{SU(1,1)}\ket{tm_t}=t(t-1)\ket{tm_t}.
\end{equation}
If $t$ is an integer or half integer, then the irrep is discrete.  If $t>0$ then the irrep is bounded below by  $t$, and if $t<0$ then the irrep is bounded above by $t$.  In other words,
\begin{equation}
m_t=\begin{cases}t,t+1,...&t>0\\
-t,-t-1,...&t<0\end{cases},\label{tweights}
\end{equation}
as shown in Fig.~\ref{fig-weights}(b) for $t>0$.


\subsection{Representation of SU(1,1) on the Laguerre radial basis \label{alg}} 
In this section, we construct a representation of the Lie algebra of
the radial group, SU(1,1), using the $S_{nl}(r)$ radial functions as
the basis of the vector space.  Recall from \eqref{eqn-Snl-eigen} that the
$S_{nl}$ basis is generated by varying $\alpha_{nl}$.  We rearrange
\eqref{eqn-Snl-eigen} to rewrite it as an eigenvalue problem for
$\alpha_{nl}$:
\begin{equation}
\left(- r\frac{d^2}{d  r^2}-\frac{d}{d  r}+\frac{(l+1)^2}{ r}+ r\right)S_{nl}(r)=2\alpha_{nl}S_{nl}(r).\label{T3}
\end{equation} 
Following a similar approach to the method used in \cite{jpa-26-1993-1601-Cooper, ijqc-12-1977-875-Cizek} to construct representations of SU(1,1) for the Coulomb functions, we define
\begin{equation}
T_3=-\frac{1}{2}\left( r\frac{d^2}{d r^2}+\frac{d}{d r}-\frac{(l+1)^2}{ r}- r\right)
\end{equation}
and
\begin{equation}
T_{\pm}=-\frac{1}{2}\left( r\frac{d^2}{d r^2}+\frac{d}{d r}-\frac{(l+1)^2}{ r}+ r\mp 2 r\frac{d}{d r}\mp 1\right), 
\end{equation}
which satisfy the commutation relations in \eqref{com1} and \eqref{negative} and the relation $T_+^\dagger=T_-$.  The Casimir invariant is given by
\begin{equation}
C_{\mathrm{SU(1,1)}}=T_3^2-\frac12(T_+T_-+T_-T_+)=(l+\frac{3}{2})(l+\frac{1}{2})\label{casmir}.
\end{equation}
Thus the representation is discrete, with $t=l+3/2$ and weights
$m_t=n+l+3/2$ ($n=0,1,....$).  Note that $m_t=\alpha_{nl}$, with
$\alpha_{nl}$ as defined in~\eqref{SCSde}.  From \eqref{m} and \eqref{pm}, the actions of the
generators are expressed in terms of the labels $n$ and $l$ as
\begin{equation}
T_3S_{nl}=(n+l+\tfrac32)S_{nl}
\end{equation}
\begin{equation}
T_+S_{nl}=\sqrt{(n+1)(n+2l+3)}S_{n+1,l}
\end{equation}
\begin{equation}
 T_-S_{nl}=\sqrt{n(n+2l+2)}S_{n-1,l}.
\end{equation} 

\subsection{Deriving the action of radial operators using the SU(1,1) algebra \label{radial MEs}}
We can use the SU(1,1) algebra to easily determine the actions of $ r$, $1/r$ and $d/dr$ on the functions $S_{nl}(r)$ following a method similar to that used by Rowe in \cite{jpa-38-2005-10181-Rowe} to calculate the matrix elements for the generalized harmonic oscillator functions.  We begin with the elements of the algebra 
\begin{equation}
 r=\frac{1}{2}(2T_3-T_+-T_-) \label{el1},
\end{equation}
and
\begin{equation}
 r\frac{d}{dr}=\frac{1}{2}\left(T_+-T_--1\right)\label{el2}.
\end{equation}
Applying  the operators in \eqref{el1} and \eqref{el2} to $S_{nl}$, we immediately have that
\begin{multline}
 r S_{nl}( r)=(n+l+\frac{3}{2})S_{nl}(r)-\frac{1}{2}\sqrt{n(n+2l+2)}S_{n-1,l}(r)\\
 -\frac{1}{2}\sqrt{(n+1)(n+2l+3)}S_{n+1,1}(r)\label{r}
\end{multline}
and 
\begin{multline}
 r\frac{d}{dr}S_{nl}(r)=\frac{1}{2}\sqrt{(n+1)(n+2l+3)}S_{n+1,l}(r)\\-\frac{1}{2}\sqrt{n(n+2l+2)}S_{n-1,l}(r)-\frac{1}{2}S_{nl}(r)\label{rdr}.  
\end{multline}
We can obtain the action of the operator $1/r$, which is not an element of the SU(1,1) algebra, using a recurrence relationship that we deduce using the orthogonality of the $S_{nl}(r)$ functions.  Starting with \eqref{r}, we divide through by $r$, multiply through by $S_{ml}(r)$ and integrate to obtain
\begin{multline}
 \delta_{n,m}=(n+l+\frac{3}{2})f_{nm}-\frac{1}{2}\sqrt{n(n+2l+2)}f_{n-1,m}
-\frac{1}{2}\sqrt{(n+1)(n+2l+3)}f_{n+1,m},\label{recur}
\end{multline}
where 
\begin{equation}
f_{nm}=\int_0^\infty d r S_{ml}(r)\frac{1}{ r}S_{nl}(r).
\end{equation}
Starting with $f_{00}=1/(l+1)$, obtained via direct integration, the recurrence relationship \eqref{recur} can be solved to find that
\begin{equation}
f_{nm}=\frac{1}{l+1}\sqrt{\frac{n!(m+2l+2)!}{m!(n+2l+2)!}},\qquad n\ge m.
\end{equation}
A similar result, obtained by simply interchanging $m$ and $n$, holds for  $n<m$. From these results we have that 
\begin{multline}
 \frac{1}{ r}S_{nl}(r)=\frac{1}{l+1}\sum_{i\ge n+1}\sqrt{\frac{i!(n+2l+2)!}{n!(i+2l+2)!}}S_{il}(r)
+\frac{1}{l+1}\sum_{i\le n}\sqrt{\frac{n!(i+2l+2)!}{i!(n+2l+2)!}}S_{il}(r).\label{1/r}
\end{multline}
Finally, the action of the radial derivative can be determined by noting that $
d/dr=r^{-1}(r d/dr)$.  Combining \eqref{rdr} and \eqref{1/r}, 
\begin{equation}
 \frac{d}{dr}S_{nl}(r)=\sum_{i\ge n+1}\sqrt{\frac{i!(n+2l+2)!}{n!(i+2l+2)!}}S_{il}(r)-\sum_{i\le n-1}\sqrt{\frac{n!(i+2l+2)!}{i!(n+2l+2)!}}S_{il}(r).\label{derivative}
\end{equation}
The actions of  other operators, built from these basic operators, can
be calculated by combining   \eqref{r},   \eqref{1/r} and   \eqref{derivative}. 

Once we have the action of a radial operator $R(r)$ on $S_{nl}$, we
can extract the matrix element $\braket{n'l|R(r)|nl}$ by inspection
as the coefficient of $S_{n'l}$ in the sum, using the orthogonality
of the radial functions $S_{n,l}$.  For example, if $n'\ge n+1$, then
\begin{equation}
\begin{aligned}
\braket{n'l|\frac{d}{dr}|nl}&=\sum_{i\ge n+1}\sqrt{\frac{i!(n+2l+2)!}{n!(i+2l+2)!}}\delta_{n'i}-\sum_{i\le n-1}\sqrt{\frac{n!(i+2l+2)!}{i!(n+2l+2)!}}\delta_{n'i}\\
&=\sqrt{\frac{n'!(n+2l+2)!}{n!(n'+2l+2)!}}.\label{example}
\end{aligned}
\end{equation}
Expression for the matrix elements for $r$, $r^2$,  $1/r$, $1/ r^2$, $rd/dr$, $d/dr$ and $d^2/dr^2$   are given in Appendix~\ref{app-tables} (Table~\ref{Smatrix}).  


\section{Shift operators\label{shift sec}}

In the previous section, we derived expressions for matrix elements of
operators between Laguerre radial functions within the same SU(1,1)
irrep, \textit{i.e.}, matrix elements between the three-dimensional basis functions with the same angular quantum number $l$.  However,
many operators of interest in physics are not strictly radial, scalar
operators (\textit{e.g.}, ${\bf r}$ and ${\boldsymbol\nabla}$ are vector
operators or, equivalently, spherical tensors of rank 1) and may
therefore have nonvanishing matrix elements between states with
different $l$.  If we consider a spherical tensor operator
$\mathcal{O}_{a\alpha}$ which factorizes into radial and angular
parts, as discussed in Sec.~\ref{operator map}, then recall
from~(\ref{matrixrel}) that the problem reduces to evaluating the
radial matrix element $\braket{n'l'|\gamma(R)|nl}$, with $l'\neq l$,
that is, connecting Laguerre radial functions lying in different
SU(1,1) irreps.

The matrix elements of radial operators between radial functions in
different irreps ($l'\neq l$) can be written in terms of the matrix
elements we have already calculated, \textit{i.e.}, between members of the same
irrep ($l'=l$), provided we can obtain analytic relations allowing us
to expand the members of one of the irreps (namely, $l'$) as linear
combinations of those in the other irrep (namely, $l$).  This relation
may be obtained through shift operators, $A^\dagger$ and $A$, introduced in
Sec.~\ref{sec-intro}.

To see how we may obtain such shift operators, recall
from~(\ref{eqn-Snl-eigen}) that the Laguerre functions are eigenfunctions
of a differential operator.  In particular, the operator
\begin{equation}
\label{eqn-Hnl}
H_{nl}=-\frac{d^2}{d r^2}-\frac1r\frac{d}{d
  r}+\frac{(l+1)^2}{r^2}-\frac{2\alpha_{nl}}{r}
\end{equation}
may be taken as a Hamiltonian operator defining the Laguerre functions
(of a single irrep $l$) as its eigenfunctions, except that this
Hamiltonian involves an $n$-dependent coefficient $\alpha_{nl}$ and is
thus actually a quantum number dependent Hamiltonian (the
eigenfunctions of interest are also degenerate, with eigenvalue $-1$).

The problem of relating Laguerre functions from different irreps (of
different $l$) thus amounts to the problem of relating the
eigenfunctions of a sequence of Hamiltonians, which are defined
through discrete variation of a parameter $l$.  This is the problem
addressed by the factorization method of Infeld and
Hull~\cite{rmp-23-1951-21-Infeld} or, equivalently, for present
purposes, the factorization method used in supersymmetric quantum
mechanics (SUSY QM)~\cite{prep-251-1995-267-Cooper} for
shape invariant potentials~\cite{jetpl-38-1983-25-Gendenshtein}.

\subsection{Factorization with quantum number dependence\label{Ndep}}

The essential idea of
factorization
is that, if a Hamiltonian can be written in a factorized form
$H=\Ad \Ax$, this immediately gives rise to a partner
Hamiltonian $H'=\Ax \Ad$, with (at least some) eigenvalues which are 
degenerate to those of $H$, and with eigenfunctions which are related to those of
$H$~--- in fact, related by the action of the operators $\Ad$ and $\Ax$.
Factorization is a particularly powerful technique in problems where
we have an extended (perhaps infinite) sequence or ``tower'' of
related Hamiltonians $H_m=\Ad_m \Ax_m$  ($m=1$, $2$,
$\ldots$), where $m$ is a discrete parameter in the Hamiltonian.  Then we likewise have extended
sets of degenerate states $\psi_{nm}$ ($n=0$, $1$,
$\ldots$ and $m=1$, $2$,
$\ldots$), related to each other by
the action of the operators $\Ad_m$ and $\Ax_m$.  
SUSY QM provides a general
formalism for finding such a factorization, through the
superpotential~\cite{prep-251-1995-267-Cooper}.  However, for many
familiar exactly-solvable Schr\"odinger equation problems, the factorizations have been cataloged by Infeld
and Hull~\cite{rmp-23-1951-21-Infeld}, and the problem of finding a
factorization reduces to that of transforming the problem into one of
their standard forms.

It may be helpful to note both the analogy to the familiar
factorization of the one-dimensional harmonic oscillator Hamiltonian,
$H=a^\dagger a+1/2$, explored by Dirac~\cite{dirac1958:qm}, as well as
the differences from this factorization.  In the oscillator problem,
the operators $a^\dagger$ and $a$ are simply \textit{laddering}
operators, between eigenfunctions of the \textit{same} Hamiltonian,
serving to change the number of nodes.  However, the operators $\Ad$
and $\Ax$ in the more general factorization are \textit{shift}
operators, between eigenfunctions of \textit{different} Hamiltonians,
though in fact their action also changes the number of nodes.

While factorization is a powerful technique for solving eigenproblems,
we are interested here primarily in the shift operation which it
provides, towards our goal of relating Laguerre radial functions in
different irreps.  We therefore summarize the necessary ideas behind
the shift operation, then show how the relevant 
relations can readily be extended to accomodate quantum number
dependent Hamiltonians, such as encountered for the Laguerre functions
in~(\ref{eqn-Hnl}).  We refer the reader to the more extensive
discussions in Refs.~\cite{rmp-23-1951-21-Infeld,prep-251-1995-267-Cooper,hecht2000:qm} for a more comprehensive discussion of
 the families of potentials which lead to
factorizable Hamiltonians, the solution process, and the nature of the
eigenspectra which result.  Our notational conventions for
factorization follow those of, \textit{e.g.},
Refs.~\cite{prep-251-1995-267-Cooper,jpa-38-2005-10181-Rowe}, in the use of $\Ad$ and
$\Ax$ for the shift operators.  

To start with this simplest case, of two partner Hamiltonians, suppose
the Hamiltonian for some one-dimensional eigenproblem can be written as
\begin{equation}
H_1(r)=\Ad_{1}(r) \Ax_{1}(r)+K_1, \label{Hamiltonian1}
\end{equation}
where, in addition to the factorized portion of the Hamiltonian, it is
convenient to permit a constant term $K_1$.  (Since the primary
concern in the present work is with radial wave functions on
$\mathbb{R}^+$, we have chosen to denote the independent variable as
$r$ rather than as $x$, but the discussion here is relevant without
modification to one-dimensional problems on $\mathbb{R}$.) Then the
eigenspectrum of $H_1$ is related in a simple way to that of the
partner Hamiltonian
\begin{equation}
H_2(r)=\Ax_1(r)\Ad_1(r) +K_1\label{Hamiltonian2}.
\end{equation}
Specifically, suppose $H_1$ has eigenfunctions $\psi_{n1}$
($n=0$, $1$, $\ldots$), with associated eigenvalues $E_{n1}$, and
$H_2$ has eigenfunctions $\psi_{n2}$ ($n=0$, $1$, $\ldots$),
with associated eigenvalues $E_{n2}$, that is,
\begin{align}
H_1\psi_{n1}&=E_{n1}\psi_{n1}\label{eqn-eigen1}\\
H_2\psi_{n2}&=E_{n2}\psi_{n2}\label{eqn-eigen2}.
\end{align}

Then, simply from the form of~\eqref{Hamiltonian1}
and~\eqref{Hamiltonian2}, one can readily verify (as we shall
illustrate below in a more general context) that acting with $\Ax_1$
on any eigenfunction $\psi_{n1}$ of $H_1$ gives an eigenfunction of
$H_2$, still with eigenvalue $E_{n1}$.  Similarly, acting with $\Ad_1$
on any eigenfunction $\psi_{n2}$ of $H_2$ gives an eigenfunction of
$H_1$, still with eigenvalue $E_{n2}$.  Thus, eigenfunctions of $H_1$
and $H_2$ must occur as degenerate partners.
The exception arises if
$\Ax_1$ (or $\Ad_1$)  annihilates the wave function on which
it acts, that is, if it yields the null function instead of a \textit{bona fide}
eigenfunction of the partner Hamiltonian.  We shall restrict our
attention to the typical case (see Refs.~\cite{rmp-23-1951-21-Infeld,prep-251-1995-267-Cooper,hecht2000:qm}) in which $\Ax_1 \psi_{01}=0$.
That is, the ground state eigenfunction of $H_1$ has no partner.  Then
the remaining eigenfunctions and eigenvalues follow the correspondence
(for $n=1$, $2$, $\ldots$)
\begin{align}
\psi_{n1}(r)&\propto \Ad_1(r)\psi_{n-1,2}(r)\label{SUSYS}\\
\psi_{n-1,2}(r)&\propto \Ax_1(r)\psi_{n1}(r)\label{SUSYP}\\
E_{n-1,2}&=E_{n1}.\label{SUSYE}
\end{align}

The factorization process may, however, continue.  Suppose $H_2$, in
turn, can be refactored in some way, distinct from~\eqref{Hamiltonian2}, as
\begin{equation}
H_2(r)=\Ad_2(r) \Ax_2(r)+K_2\label{Hamiltonian2-2}.
\end{equation}
Then the same pattern can be followed, as above, to relate the eigenspectrum of $H_2$ to that of a partner Hamiltonian
\begin{equation}
H_3(r)=\Ax_2(r)\Ad_2(r)+K_2,
\end{equation}
and relations on the eigenfunctions and eigenvalues analogous to~\eqref{SUSYS}--\eqref{SUSYE} can be
obtained.

More generally, we may consider an infinite ``tower'' of
related Hamiltonians $H_m$, parametrized by a discrete index $m$
($m=1$, $2$, $\ldots$), where successive Hamiltonians $H_m$
and $H_{m+1}$ have related factorizations
\begin{align}
H_m(r)&=\Ad_{m}(r)\Ax_{m}(r)+K_{m}\label{Hamiltonianmu1}\\
H_{m+1}(r)&=\Ax_{m}(r)\Ad_{m}(r)+K_{m}\label{Hamiltonianmu2}.
\end{align}
We now have a sequence of related eigenproblems
\begin{equation}
H_m(r)\psi_{nm}(r)=E_{nm}\psi_{nm}(r)\label{eqn-eigen-m}.
\end{equation}
Their eigenfunctions and eigenvalues are related by
\begin{align}
\psi_{nm}(r)&\propto \Ad_{m}(r)\psi_{n-1,m+1}(r)\label{rel1}\\
\psi_{n-1,m+1}(r)&\propto \Ax_{m}(r)\psi_{nm}(r)\label{rel2}\\
E_{n-1,m+1}&=E_{nm}\label{rel3},
\end{align}
etc.  The identification of $n$ quantum numbers ($n=0$, $1$, $\ldots$)
across the different eigenproblems here again rests on the further
conventional condition that the ground state for each $H_m$ be annihilated by the
$\Ax$ operator, \textit{i.e.}, $\Ax_{m}\psi_{0m}=0$ or, equivalently, as may
be verified, $E_{0m}=K_m$.

\begin{figure}
\begin{center}
\includegraphics[width=0.8\hsize]{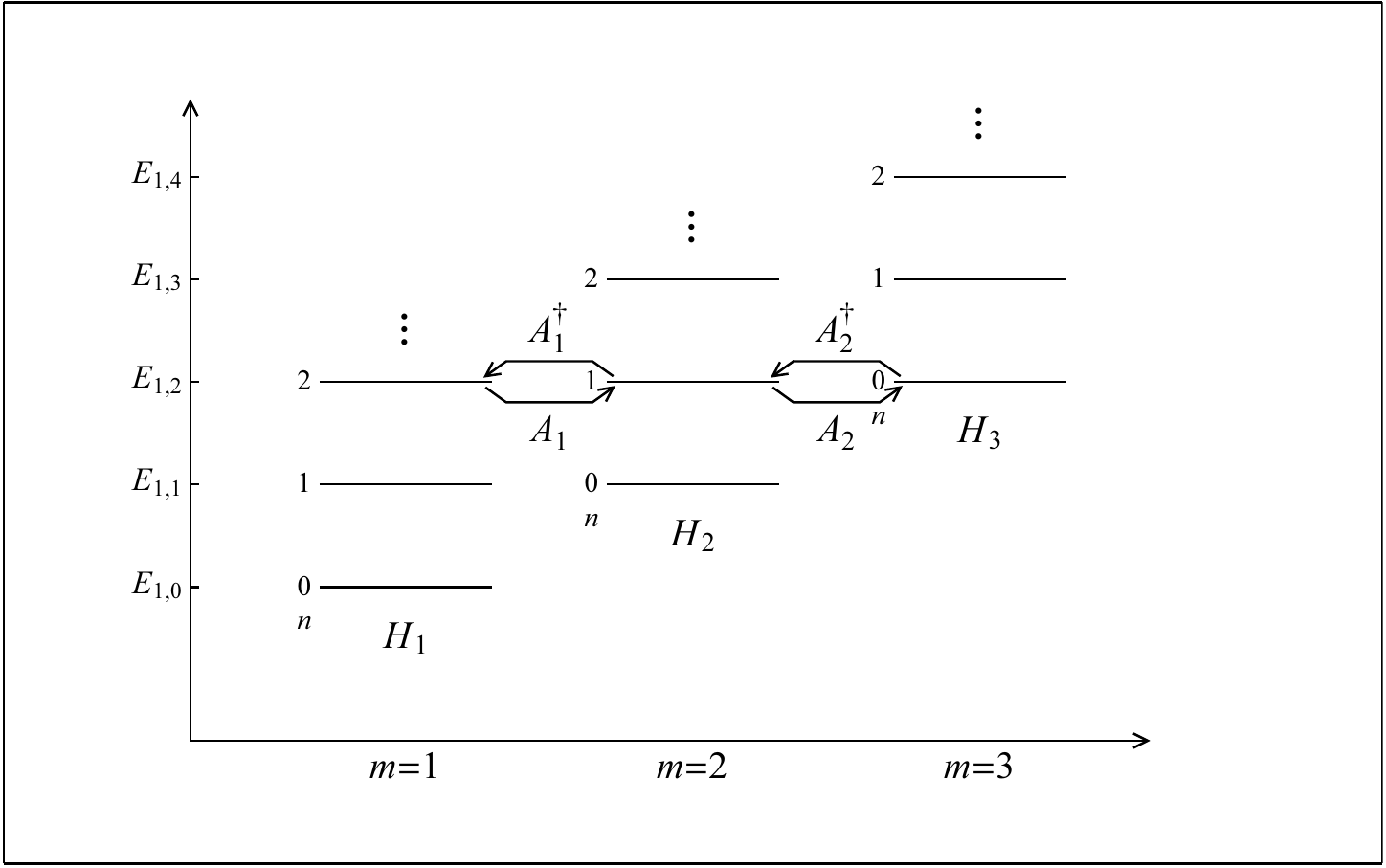}
\end{center}
\caption{Structure of the relations among eigenfunctions and eigenvalues for a sequence of factorized Hamiltonians $H_m$, defined in~\eqref{rel1}--\eqref{rel3}.}
\label{fig-lattice-susy}
\end{figure}

Let us take a moment to interpret the energy and operator relations
indicated in~\eqref{rel1}--\eqref{rel3}, which are illustrated
schematically in Fig.~\ref{fig-lattice-susy}.  From~\eqref{rel1}, it
is seen that the operator $\Ad_m$ acts on the wave function
$\psi_{n-1,m+1}$ to give the wave function $\psi_{nm}$.  It thus
\textit{raises} the quantum number $n$, which counts the number of
nodes in the wave function, and is therefore reminiscent of Dirac's
$a^\dagger$ operator.  However, the initial and final wave functions
are eigenfunctions of \textit{different} Hamiltonians, as the
Hamiltonian index has simultaneously been \textit{lowered}.  Thus,
$\Ad_m$ is not merely a \textit{raising} operator, but rather a
\textit{shift} operator.  Similarly, from~\eqref{rel2}, it is seen
that $\Ax_m$ is likewise a shift operator, but with the inverse action
on the node number and Hamiltonian indices.  Finally,
from~\eqref{rel3}, the eigenvalues form degenerate multiplets across
the different Hamiltonians, with
$E_{n0}=E_{n-1,1}=E_{n-2,2}=\ldots$. In fact, by repeated application
of \eqref{rel1}--\eqref{rel3}, the eigenfunctions and eigenvalues of
any two Hamiltonians in the sequence can be related.  Comparing the
shift relations among eigenspectra arising from factorization
(Fig.~\ref{fig-lattice-susy}) with those anticipated for shift
operators between $\grpsu{1,1}$ irreps (Fig.~\ref{fig-lattice-su11}), we may
already see the essential resemblance.

We may now generalize these relations to the case in which the
Hamiltonians are $n$-dependent, that is, to eigenproblems of the form
\begin{equation}
H_{nm}(r)\psi_{nm}(r)=E_{nm}\psi_{nm}(r)\label{eqn-eigen-mn}.
\end{equation}
The factorizations~\eqref{Hamiltonianmu1} and~\eqref{Hamiltonianmu2}
then generalize to
\begin{align}
H_{nm}(r)&=\Ad_{nm}(r)\Ax_{nm}(r)+K_{nm}\label{NHam1}\\
H_{n-1,m+1}(r)&=\Ax_{nm}(r)\Ad_{nm}(r)+K_{nm},\label{NHam2},
\end{align}
again with the assumption that the ground state eigenfunctions are
annihilated, \textit{i.e.}, $\Ax_{0m}(r)\psi_{0m}(r)=0$.
We label the latter Hamiltonian with the index $n-1$, rather than $n$, in recognition of the
anticipated correspondence of eigenfunctions across the Hamiltonians. 

We may now obtain $n$-dependent shift operator relations
\begin{align}
\psi_{nm}(r)&\propto \Ad_{nm}(r)\psi_{n-1,m+1}(r)\label{NSUSY1}\\
\psi_{n-1,m+1}(r)&\propto \Ax_{nm}(r)\psi_{nm}(r),\label{NSUSY2}
\end{align}
while the eigenvalue relation $E_{n-1,m+1}=E_{nm}$ from~\eqref{rel3}
still holds.
These relations follow by first establishing that $\Ad_{nm}$ acts on the eigenfunction
$\psi_{n-1,m+1}$ of
$H_{n-1,m+1}$ to give an eigenfunction of $H_{nm}$, with the same
eigenvalue $E_{n-1,m+1}$:
\begin{equation}
\label{N3}
\begin{aligned}
H_{nm}[\Ad_{nm}(r)\psi_{n-1,m+1}(r)]
&=[\Ad_{nm}(r)\Ax_{nm}(r)+K_{nm}]\Ad_{nm}(r)\psi_{n-1,m+1}(r)\\
&=\Ad_{nm}(r)[\Ax_{nm}(r)\Ad_{nm}(r)+K_{nm}]\psi_{n-1,m+1}(r)\\
&=\Ad_{nm}(r)H_{n-1,m+1}\psi_{n-1,m+1}(r)\\
&=E_{n-1,m+1}[\Ad_{nm}(r)\psi_{n-1,m+1}(r)],
\end{aligned}
\end{equation}
where we have successively applied the definition of $H_{nm}$
from~\eqref{NHam1}, reassociated, recognized the definition of $H_{n-1,m+1}$
from~\eqref{NHam2}, and applied the eigenproblem definition from~\eqref{eqn-eigen-mn}.
It then may be verified, by a similar argument, that $\Ax_{nm}$ acts on the eigenfunction
$\psi_{nm}$ of
$H_{nm}$ to give an eigenfunction of $H_{n-1,m+1}$, with the same
eigenvalue $E_{nm}$, \textit{i.e.},
\begin{equation}
\label{M1}
H_{n-1,m+1}[\Ax_{nm}(r)\psi_{nm}(r)]
=E_{nm}[\Ax_{nm}(r)\psi_{nm}(r)].
\end{equation}

To apply the shift operators in practice, it is desirable not only to
know that a proportionality exists, as
in~\eqref{NSUSY1} and~\eqref{NSUSY2}, but also to know the constant of
proportionality, so that we may construct \textit{normalized shift operators} $\Adn_{nm}$ and $\Axn_{nm}$.
We require that
these operators, when  acting on normalized eigenfunctions, then yield
normalized eigenfunctions, so that the proportionalities~\eqref{NSUSY1} and~\eqref{NSUSY2} become equalities
\begin{align}
\psi_{nm}(r)&= \Adn_{nm}(r)\psi_{n-1,m+1}(r)\label{NSUSY1-norm}\\
\psi_{n-1,m+1}(r)&= \Axn_{nm}(r)\psi_{nm}(r).\label{NSUSY2-norm}.
\end{align}

The constants of proportionality in~\eqref{NSUSY1} and~\eqref{NSUSY2}
follow uniquely, within phase, from the assumed
adjoint relationship between the operators $\Ad_{nm}$ and $\Ax_{nm}$
and from the requirement that the wave functions $\psi_{nm}$ be
normalized.
In particular, the norm of $\Ad_{nm}\psi_{n-1,m+1}$ is readily evaluated,
since (we switch to bracket notation to facilitate representing the inner product)
\begin{equation}
\begin{aligned}
\lvert \Ad_{nm}\psi_{n-1,m+1} \rvert^2 
&=\braket{\psi_{n-1,m+1}|\Ax_{nm}\Ad_{nm}|\psi_{n-1,m+1}}\\
&=\braket{\psi_{n-1,m+1}|\bigl( H_{n-1,m+1}  -K_{nm}\bigr)|\psi_{n-1,m+1}}\\
&=E_{n-1,m+1} -K_{nm},
\end{aligned}
\end{equation}
where we have recognized the relationship between $\Ax_{nm}\Ad_{nm}$
and $H_{n-1,m+1}$ by~\eqref{NHam2} and then applied the eigenproblem definition from~\eqref{eqn-eigen-mn}.
The result may simply be rewritten as $\lvert \Ad_{nm}\psi_{n-1,m+1} \rvert^2 = E_{nm} -K_{nm}$ by the eigenvalue relation~\eqref{rel3}.
Similarly,
considering the norm of $\Ax_{nm}\psi_{nm}$ gives an identical value  $\lvert \Ax_{nm}\psi_{nm} \rvert^2 = E_{nm} -K_{nm}$.

Specifically, to construct the normalized shift operators, let
\begin{align}
\Adn_{nm} = \sigma_{nm} c_{nm} \Ad_{nm}\label{normalized1}\\
\Axn_{nm} = \sigma'_{nm} c'_{nm} \Ax_{nm}\label{normalized2},
\end{align}
where $\sigma_{nm}$ and $\sigma'_{nm}$ represent the phases, and
$c_{nm}$ and $c'_{nm}$ represent the normalization factors.  
Thus, insisting that both
$\psi_{nm}=\Adn_{nm}\psi_{n-1,m+1}$ and $\psi_{n-1,m+1}=\Axn_{nm}\psi_{nm}$ be
normalized gives $c_{nm}=c'_{nm}=(E_{nm}-K_{nm})^{-1/2}$.

The phases (signs) of the solutions to an
eigenproblem~\eqref{eqn-eigen-m} or~\eqref{eqn-eigen-mn} are not
determined by the eigenproblem itself, but rather are a matter of
convention.  For instance, it may be chosen that the wave functions
are positive as $r\rightarrow0$ or positive as $r\rightarrow\infty$.
Thus, the phases $\sigma_{nm}$ and $\sigma'_{nm}$ appearing in the
normalized shift operators cannot be
deduced purely from the factorization formalism but must rather be
chosen so that the shift operators yield wave functions with phases
consistent with the adopted convention.  The self-consistency requirement
that $\psi_{nm}=\Adn_{nm}\psi_{n-1,m+1}=\Adn_{nm}\Axn_{nm}\psi_{nm}$
imposes the constraint that $\sigma_{nm}=\sigma'_{nm}$.

If the normalized shift operators are known for all $m$, then we have
a way of expressing eigenfunctions of any Hamiltonian $H_{m+k}$ in
terms of those of $H_m$, via~\eqref{NSUSY1-norm}
or~\eqref{NSUSY2-norm}.  In particular, if $k$ is a positive integer,
then we repeatedly apply $\Axn$ to obtain
\begin{equation}
\psi_{n,m+k} = \Axn_{n+1,m+k-1} \Axn_{n+2,m+k-2} \cdots \Axn_{n+k,m}  \psi_{n+k,m}.
\end{equation}

\subsection{Shift operators for Laguerre functions\label{CSshift}}

We now deduce shift operators on the $l$
quantum number of the Laguerre radial functions $S_{nl}(r)$.  We deduce these shift
operators by recasting the differential equation \eqref{eqn-Snl-eigen}
into the form of a factorizable $n$-dependent Hamiltonian
eigenproblem, so that we may use the formalism
laid out in Sec.~\ref{Ndep}.  Specifically, \eqref{eqn-Snl-eigen} can
be rewritten in terms of the Hamiltonian operator \eqref{eqn-Hnl} as 
\begin{equation}
H_{nl}S_{nl}(r)=-1S_{nl}(r), \label{S_Hamiltonian_eq}
\end{equation}
where $E_{nl}=-1$ for all $n$ and $l$.  However, to put this
Hamiltonian into one of the standard factorizable forms cataloged by
Infeld and Hull~\cite{rmp-23-1951-21-Infeld}, we must first eliminate
the first-derivative term, through a similarity transformation
$r^{1/2}H_{nl}r^{-1/2}$.  The
Hamiltonian equation \eqref{S_Hamiltonian_eq} becomes
\begin{equation}
\left[-\frac{d^2}{d r^2}+\frac{(l+\frac12)(l+\frac32)}{ r^2}-\frac{2\alpha_{nl}}{r}\right]R_{nl}(r)=-1R_{nl}(r),\label{trans dif}
\end{equation}
where 
\begin{equation}
R_{nl}(r)=r^{1/2}S_{nl}(r).  
\end{equation}          
Comparing \eqref{trans dif} with the standard form for Coulomb-type
problems, given by~(8.0.1) of Ref.~\cite{rmp-23-1951-21-Infeld}, yields the factorization~\cite{fn-parameter-identification}
\begin{equation}
\label{eqn-factorize-laguerre-shift-ops}
\tilde{A}^\dagger_{nl}(r)=\left(-\frac{d}{d r}-\frac{l+\frac32}{r}+\frac{\alpha_{nl}}{l+\frac32}\right)
\qquad \tilde{A}_{nl}(r)=\left(\frac{d}{d r}-\frac{l+\frac32}{r}+\frac{\alpha_{nl}}{l+\frac32}\right)
\end{equation}
and constant term
\begin{equation}
\label{eqn-factorize-laguerre-shift-constant}
 K_{nl}=-\frac{\alpha_{nl}^2}{(l+\frac32)^2}.
\end{equation}
The tildes on $\tilde{A}^\dagger$ and $\tilde{A}$ indicate that these
are the shift operators corresponding to the similarity transformed
Hamiltonian operator \eqref{trans dif}.  We also note that
$K_{0l}=-1=E_{0l}$ ($l=0,1,\ldots$), thus satisfying the condition to obtain the relationships \eqref{rel1}--\eqref{rel3} between partner
Hamiltonian eigenfunctions and eigenvalues, equivalent to
$\tilde{A}_{0l}R_{0l}=0$ and thus $A_{0l}S_{0l}=0$ (see
Sec. \ref{Ndep}).

From \eqref{eqn-factorize-laguerre-shift-ops} and \eqref{eqn-factorize-laguerre-shift-constant}, we can write \eqref{trans dif} in the $n$-dependent
factorized form of~\eqref{NHam1}, as 
\begin{multline}
\left[\left( -\frac{d}{d r}-\frac{l+\frac32}{r}+\frac{\alpha_{nl}}{l+\frac32}\right)\left( \frac{d}{d r}-\frac{l+\frac32}{r}+\frac{\alpha_{nl}}{l+\frac32}\right)-\frac{\alpha_{nl}^2}{(l+\frac32)^2}\right]R_{nl}(r)
=-1R_{nl}(r).\label{Ham}
\end{multline}
  The normalized shift operators
  \eqref{normalized1}--\eqref{normalized2} for the functions $R_{nl}(r)$ are given by
\begin{align}
 \tilde{\mathcal{A}}_{nl}^\dagger(r)=\frac{-(l+\frac32)}{\sqrt{n(n+2l+3)}}\left(- \frac{d}{d r}-\frac{l+\frac32}{r}+\frac{\alpha_{nl}}{l+\frac32}\right)  
  \label{norm R shift1}\\ 
  \tilde{\mathcal{A}}_{nl}(r)=\frac{-(l+\frac32)}{\sqrt{n(n+2l+3)}}\left(\frac{d}{d r}-\frac{l+\frac32}{r}+\frac{\alpha_{nl}}{l+\frac32}\right).  \label{norm R shift2}
\end{align}
Here we adopt phases $\sigma_{nl}=-1$, to enforce the convention that $R_{nl}(r)$ [and thus
$S_{nl}(r)$ below] is always positive at the origin.

To obtain the shift
operators for the functions $S_{nl}(r)$, we must apply the inverse similarity
transformation.  Thus, $\Adn_{nl}(r)=r^{-1/2}\Adnt_{nl}(r) r^{1/2}$ gives
\begin{equation}
\mathcal{A}^{\dagger}_{nl}(r)=\frac{-(l+\frac32)}{\sqrt{n(n+2l+3)}}\left(-\frac{d}{d r}-\frac{l+2}{r}+\frac{\alpha_{nl}}{l+\frac32}\right),
\end{equation}
and, similarly, $\Axn_{nl}(r)=r^{-1/2}\Axnt_{nl}(r) r^{1/2}$ gives
\begin{equation}
\mathcal{A}_{nl}(r)=\frac{-(l+\frac32)}{\sqrt{n(n+2l+3)}}\left(\frac{d}{d r}-\frac{l+1}{r}+\frac{\alpha_{nl}}{l+\frac32}\right).
\end{equation}
\subsection{Deriving the action of radial operators across irreps
  using the shift operators}
\label{CSshiftme}

Now that we have shift operators relating $S_{nl}(r)$ functions with different $l$, 
\begin{equation}
S_{n,l+1}(r)=\mathcal{A}_{n+1,l}(r)S_{n+1,l}(r)\qquad
S_{nl}(r)=\mathcal{A}_{nl}^\dagger(r)S_{n-1,l+1}(r), \label{shiftrel}
\end{equation}
we can derive expressions for the matrix elements between functions
$S_{nl}$ and $S_{n'l'}$ for $l'=l\pm k$, for $k=1,2,\ldots$, by
writing $S_{n',l+k}$ in terms of $S_{nl}$ using the
relationships \eqref{shiftrel}.  Note that the matrix elements for
$l'=l-k$ can be obtained from the matrix elements of $l'=l+k$ by
making the appropriate substitutions (Appendix~\ref{app-tables}), and thus it is sufficient to
consider only $l'=l+k$.

For $l'=l\pm1$, we simply apply the raising shift operator to $S_{n+1,l}$
\begin{equation}
\begin{aligned}
S_{n,l+1}(r)&=\mathcal{A}_{n+1,l}(r)S_{n+1,l}(r)\\
&=\frac{-(l+\frac32)}{\sqrt{(n+1)(n+2l+4)}}\left(\frac{d}{d r}-\frac{l+1}{r}+\frac{\alpha_{nl}}{\alpha_{0l}}\right)S_{n+1,l}(r)\label{l+1}.
\end{aligned}
\end{equation}
Applying   \eqref{1/r} and   \eqref{derivative} to \eqref{l+1},
\begin{multline}
S_{n,l+1}(r)=\frac{-(l+\frac32)}{\sqrt{(n+1)(n+2l+4)}}\left[\frac{n+1}{l+\frac32}S_{n+1,l}(r)\right.\\
-\left.2\sum_{i=0}^n\left(\frac{(n+1)!(i+2l+2)!}{i!(n+2l+3)!}\right)^{1/2}S_{il}(r)\right].
\end{multline}

For $l'=l+2$, the raising shift operator must be applied twice,
\begin{equation}
S_{n,l+2}(r)=\mathcal{A}_{n+1,l+1}(r)S_{n+1,l+1}(r)=\mathcal{A}_{n+1,l+1}(r)\mathcal{A}_{n+2,l}(r)S_{n+2,l}(r).
\end{equation}
From this we have that
\begin{equation}
 S_{n,l+2}(r)=\sqrt{\frac{(n+1)(n+2)}{(n+2l+6)(n+2l+5)}}S_{n+2,l}(r)
+\sum_{i=0}^{n+1}c_{i}\sqrt{\frac{n!(i+2l+2)!}{i!(n+2l+6)!}}S_{il}(r)\nonumber,
\end{equation}
where
\begin{equation}
c_{i}=(2l+4)(2l+3)n-(2l+5)(2l+4)i+(2l+4)(2l+3).
\end{equation}
Similarly, we could obtain the explicit relationship between
$S_{n',l+k}$ and $S_{nl}$ for any $k$ by applying the shift operator
to $S_{n-k,l}$ $k$ times.  However, we only explicitly calculate
examples of
matrix elements for $l'=l\pm1$ and $l'=l\pm2$ in this paper.

Once we have an expression for $S_{n,l+k}(r)$ in terms of $S_{nl}(r)$, we can then use the matrix elements already calculated for $l=l'$ to derive the matrix elements for $l'\ne l$.  For example, 
\begin{multline}
rS_{n,l+1}(r)=\frac{-(l+\frac32)}{\sqrt{(n+1)(n+2l+4)}}\left[\frac{n+1}{l+\frac32}rS_{n+1,l}(r)\right.\\
-\left.2\sum_{i=0}^n\left(\frac{(n+1)!(i+2l+2)!}{i!(n+2l+3)!}\right)^{1/2}rS_{il}(r)\right].
\end{multline}
From   \eqref{r}, we then deduce that
\begin{multline}
 rS_{n,l+1}(r)=-\frac{1}{2}\sqrt{(n+1)(n+2)}S_{n+2,l}+\sqrt{(n+1)(n+2l+4)}S_{n+1,l}(r)\\
-\frac{1}{2}\sqrt{(n+2l+3)(n+2l+4)}S_{nl}(r).
\end{multline}
Using the orthonormality of the $S_{nl}(r)$ we can read off the matrix elements $\braket{n'l|r|n,l+1}$ from this expression, \textit{i.e.}, 
\begin{align}
\braket{n+2,l|r|n,l+1}&=-\frac{1}{2}\sqrt{(n+1)(n+2)}\\
\braket{n+1,l|r|n,l+1}&=\sqrt{(n+1)(n+2l+4)}\\
\braket{nl|r|n,l+1}&=-\frac{1}{2}\sqrt{(n+2l+3)(n+2l+4)}
\end{align}
and all other matrix elements are zero.  Matrix elements of select operators between functions with $l'=l\pm1$ or $l'=l\pm2$ are given in Appendix~\ref{app-tables} (Tables \ref{Table l+1} and \ref{Table l+2}).   

\section{Matrix elements in the Laguerre function basis\label{lambda}} 
Now that we have a way to calculate matrix elements of radial
operators with respect to the $S_{nl}$ basis on $\mathbb{R}^+$, we can
calculate the matrix elements for irreducible tensor operators in the
$\Lambda_{nlm}$ basis on $\mathbb{R}^3$. In Sec.~\ref{operator map} we
saw that the matrix element for
$\mathcal{O}_{a\alpha}=R\Theta_{a\alpha}$ can be written~\eqref{matrixrel} as
\begin{equation}
\braket{n'l'm'|\mathcal{O}_{a\alpha}|nlm}=\braket{n'l'|\gamma(R)|nl}\braket{l'm'|\Theta_{a\alpha}|lm},
\end{equation}
where $\gamma(R)$ is given by \eqref{symtrans}.  From standard angular
momentum theory for spherical tensor operators
\cite{varshalovich1988:am}, matrix elements involving different
angular momentum projection quantum numbers $m$ and $m'$ are all
related via the Wigner-Eckart theorem, so we need only consider the
reduced matrix elements.   The reduced matrix element of
$\mathcal{O}_{a\alpha}$ is given by the product of the SO(3) reduced
matrix element of the angular operator and the matrix element of the
radial part in the $S_{nl}$ basis,
\begin{equation}
\braket{n'l'||\mathcal{O}_{a}||nl}=\braket{n'l'|\gamma(R)|nl}\braket{l'||\Theta_{a}||l}.\label{WE}
\end{equation}
Note that we follow the normalization and phase
  convention of, \textit{e.g.}, Refs.~\cite{edmonds1960:am,varshalovich1988:am} for the reduced matrix elements in the
  Wigner-Eckart theorem, \textit{i.e.},
\begin{equation}
\braket{l'm'|\Theta_{a\alpha}|lm}=\hat{l}^{-1}(lm;a\alpha|l'm')\braket{l'||\Theta_a||l},
\end{equation}
where the quantity delimited by parentheses is a Clebsch-Gordan
coefficient, and $\hat{l}=\sqrt{2l+1}$.

To illustrate this method of calculating the matrix elements in the $\Lambda_{nlm}({\bf r})$ basis, we will consider several examples, specifically, $r^k$ for $k\in \mathbb{Z}$, $\nabla^2$ and $\nabla_1$.  
\subsection{Matrix elements of the purely radial operator $r^k$}
Matrix elements of polynomial operators that depend only on $r$ are particularly simple to calculate.  In spherical tensor form, $r^k\equiv r^k{\bf 1}_{00}$.  Applying the factorization in \eqref{WE}, we can write 
\begin{equation}
\braket{n'l'||r^k{\bf 1}_{00}||nl}=\braket{n'l'|\gamma(r^k)|nl}\braket{l'||{\bf 1}_{00}||l}\label{rad1}.
\end{equation}
The reduced matrix element of the identity operator is $\braket{l'||{\bf 1}_{00}||l}=\hat{l}\delta_{l'l}$, and, 
under the transformation \eqref{symtrans}, $\gamma(r^k)=r^k$ for any $k$.  Therefore, \eqref{rad1} reduces to 
\begin{equation}
\braket{n'l'||r^k{\bf 1}_{00}||nl}=\braket{n'l|r^k|nl}\hat{l}.
\end{equation}
The radial matrix elements are already known from Sec.~\ref{radial MEs} and can be obtained from Table \ref{Smatrix} for select values of $k$.  
\subsection{Matrix elements of the scalar operator $\nabla^2$}
 The Laplacian $\nabla^2$ has the well known form \cite{varshalovich1988:am}
\begin{equation}
\nabla^2=\frac{1}{r^2}\frac{\partial}{\partial r}r^2\frac{\partial}{\partial r}-\frac{L^2}{r^2}.
\end{equation}
We factor each of the terms contributing to the matrix element of this operator into radial and angular parts, according to \eqref{WE}, and recall $L^2\ket{lm}=l(l+1)\ket{lm},$ yielding 
\begin{align}
&\braket{n'l'||\nabla^2||nl}\nonumber\\
&~~=\braket{n'l|\gamma\left(r^{-2}\frac{d}{d r}r^2\frac{d}{d r} \right)|nl}\hat{l}\delta_{ll'}-\braket{n'l'|\gamma(r^{-2})|nl}\braket{l'||L^2||l}\nonumber\\
&~~=\braket{n'l|\frac{d^2}{d r^2}|nl}\hat{l}-\braket{n'l|r^{-2}|nl}l(l+1)\hat{l}
\end{align}
The matrix elements of $d^2/d r^2$ and $r^{-2}$ can be read from \autoref{Smatrix}, and we find that 
 \begin{equation}
 \braket{n'l||\nabla^2||nl}=\begin{cases}-\frac{4n+2l+3}{2l+3}\hat{l}&n'=n\\
\frac{4n'+4l+6}{2l+3}\hat{l} \left(\frac{n!(n'+2l+2)!}{n'!(n+2l+2)!}\right)^{1/2}&n'\le n-1\\
\frac{4n+4l+6}{2l+3}\hat{l}\left(\frac{n'!(n+2l+2)!}{n!(n'+2l+2)!}\right)^{1/2}&n'\ge n+1.
 \end{cases}
 \end{equation}
 \subsection{Matrix elements of the tensor operator $\nabla_1$ \label{vec(nabla)}}
We now want to calculate the reduced matrix elements the gradient operator $\vec{\nabla}$, which can be expressed as a rank 1 tensor ${\nabla_{1}}$ with components \cite{varshalovich1988:am}
\begin{equation}
\nabla_{1\mu}=\left(\frac{4\pi}{3}\right)^{\frac12}\left\{Y_{1\mu}\frac{\partial}{\partial r}-\frac{\sqrt{2}}{r}\left[Y_1\otimes L_1\right]_{1\mu}\right\},
\end{equation}
where $Y_1$ is the irreducible tensor whose components are the spherical harmonics $Y_{1\mu}(\hat{r})$, and $L_1$ is the orbital angular momentum operator, likewise considered as a spherical tensor.   The reduced matrix element of $\nabla_1$ is thus given by
\begin{equation}
\braket{n'l'|| \nabla_1||nl}=\left(\frac{4\pi}{3}\right)^{\frac12}\braket{n'l'||\left[Y_{1}\frac{\partial}{\partial r}-\frac{\sqrt{2}}{r}\left[Y_1\otimes L_1\right]_{1}\right]||nl}\label{matrix1}.
\end{equation}
In each term, the radial and angular functions can be separated by applying \eqref{WE}.  The angular part of the second term can be factored as
\begin{multline}
\braket{n'l'||\left[Y_1\otimes L_1\right]_{1}||nl}=(-1)^{l+l'-1}\sqrt{\frac{3}{2l'+1}}\sum_{l_1}\left\{\begin{array}{ccc}1&1&1\\l&l'&l_1\end{array}\right\}\\\times\braket{n'l'||Y_1||n_1l_1}\braket{n_1l_1||L_1||nl}
\end{multline}
by the standard reduction formula for reduced matrix elements of
product operators on a single space \cite{varshalovich1988:am}.  Using
the reduced matrix elements of $Y_1$ and $L_1$ given in
\cite{varshalovich1988:am}, the expression for the reduced matrix
element of $\nabla_1$ becomes
\begin{align}
\braket{n'l'||\nabla_1||nl}
&=\sqrt{\frac{l+1}{2l+3}}\braket{n'l'|\left(\frac{d}{d r}-\frac{l+1}{r}\right)|nl}\delta_{l',l+1}\nonumber\\&\quad-\sqrt{\frac{l}{2l-1}}\braket{n'l'|\left(\frac{d}{d r}+\frac{l}{r}\right)|nl}\delta_{l',l-1}.
\end{align}
The terms contributing to the radial matrix elements can then be read from Table \ref{Table l+1}, and we have 
\begin{multline}
\braket{n',l+1||\nabla_1||nl}
=\begin{cases}-2(n'+l+3/2)\left(\frac{l+1}{2l+3}\right)^{\frac12}\left(\frac{n'!(n+2l+2)!}{n!(n'+2l+4)!}\right)^{\frac12}&n'\ge n\\
\left(\frac{n(l+1)}{(2l+3)(n+2l+3)}\right)^{1/2}&n'=n-1\\
0&n'\le n-2.\end{cases}
\end{multline}  
and 
\begin{multline}
\braket{n',l-1||\nabla_1||nl}
=\begin{cases}0&n'\ge n+2\\-\left(\frac{(n+1)l}{(2l-1)(n+2l+2)}\right)^{1/2}&n'=n+1\\-2(n'+l+1/2)\left(\frac{l}{2l-1}\right)^{\frac12}\left(\frac{n!(n'+2l)!}{n'!(n+2l+2)!}\right)^{\frac12}&n'\le n.
\end{cases}
\end{multline}

\section{Conclusion}
It has long been recognized that an $\grpsu{1,1}\times\grpso{3}$ algebraic structure, together with
shift operators between $\grpsu{1,1}$ irreps, form a
successful combination in the treatment of the classic sets of central
force eigenfunctions~--- those of the harmonic oscillator and Coulomb
problems.  We have now demonstrated that a similar framework can be
developed and effectively applied to the Laguerre functions, a basis
set of practical value in quantum mechanical one-body, few-body, and
many-body problems.

The essential results thus obtained provide for the
evaluation of matrix elements of spherical tensor operators.  This is
the immediate application of interest in formulating quantum
mechanical problems for numerical solution.  The evaluation of matrix
elements is accomplished by decomposing the full operator, in the
three-dimensional coordinate space, into parts which act only radial
or angular coordinates.  Then, recall the structure outlined in
Fig.~\ref{fig-lattice-su11}: (1)~the $\grpsu{1,1}$ algebra of the
radial functions is used to deduce radial matrix elements, while
(2)~the shift operators connect the different families of radial
functions, or $\grpsu{1,1}$ irreps, associated with different angular
functions, and (3)~the $\grpso{3}$ angular momentum algebra is used to address
the action of the operator on the angular functions themselves.  The
$\grpsu{1,1}$ algebra follows directly from the eigenvalue equation satified by
the radial functions.  The ladder operators are obtained by factorization
(or, equivalently, SUSY QM) methods, reformulated to allow for
quantum number dependence of the shift operators.  And the $\grpso{3}$
algebra provides the familiar tools of the Wigner-Eckart theorem and
Racah's reduction formulas for the angular operators.

Although the results, as presented, have been specialized to the
Laguerre functions on $\bbR^3$, their applicability is broader.  In
particular, the results for the radial functions as a basis on
$\bbR^+$ stand alone or in conjunction with any decomposition of
$N$-dimensional coordinates on $\bbR^N$ into a radial coordinate on
$\bbR^+$ and angular coordinates on the $(N-1)$-sphere $S^{N-1}$.  For
instance, functions of the Laguerre type considered here are applied
as hyperradial basis functions in few-body calculations in Jacobi
coordinates (\textit{e.g.}, Ref.~\cite{prc-86-2012-034321-Bacca}).
Alternatively, much as the $\grpsu{1,1}$ treatment of the harmonic
oscillator radial functions may be extended to a broader family of
\textit{modified harmonic oscillator functions}, by transitioning from
a discrete $\grpsu{1,1}$ irrep label $t=l+3/2$ to a continuous
label~\cite{jpa-38-2005-10181-Rowe} (physically corresponding from
the transition from a pure oscillator potential to a Davidson
potential~\cite{prsla-135-1932-459-Davidson}, containing an
additional $1/r^2$ dependence), a similar generalization may be
carried out for the present results, yielding an algebraic description
of families of modified Laguerre functions.  The general framework may
also be carried over to the Coulomb-Sturmian functions, for which the
$\grpsu{1,1}$ radial algebra is considered in
Ref.~\cite{jmp-39-1998-5811-Levai}.

Finally, it is worth noting that, so far, we have focused exclusively
on the $\grpsu{1,1}$ character of the basis functions, rather than on
the $\grpsu{1,1}$ tensorial properties of operators on the radial
space.  Through these properties, it is possible to benefit from the
Wigner-Eckart theorem, not only of the $\grpso{3}$ angular algebra as
we have done so far, but also of the $\grpsu{1,1}$ radial algebra.
Matrix elements involving different members of the same irrep
(\textit{i.e.}, different radial functions for the same angular
momentum) may then be related to each other through $\grpsu{1,1}$
Clebsch-Gordan coefficients.  The $\grpsu{1,1}$ Wigner-Eckhart
theorem, applied in this fashion, has been a powerful tool in the
construction of closed-form expressions for matrix elements in the
classic harmonic oscillator and Coulomb
eigenproblems~\cite{wybourne1974:groups,jmp-12-1971-1780-Quesne}.

\section*{Acknowledgements}
We would like to thank C.~Kolda and J.~de~Blas for valuable
discussions, M.~Wolf for comments on the manuscript and A.~I.~Rakoski
for verification of expressions for matrix elements.  This work was
supported by the Research Corporation for Science Advancement under a
Cottrell Scholar Award and by the US Department of Energy under Grant
No.~DE-FG02-95ER-40934.


\appendix

\newpage
\section{Matrix elements between Laguerre radial functions\label{app-tables}}
In the following tables, we have summarized the matrix elements of 
some of the most basic radial operators that can
be constructed from $r$, $r^{-1}$ and $d/dr$, taken in the Laguerre function radial basis.
Matrix elements within a single $\grpsu{1,1}$ irrep, \textit{i.e.},
between radial functions $S_{nl}$ of the same $l$, are given in
Table~\ref{Smatrix}.  Matrix elements between radial
functions belonging to different $\grpsu{1,1}$ irreps, corresponding
to angular momenta differing by $\Delta l=1$ and $2$, are given in
Tables~\ref{Table l+1} and~\ref{Table l+2}, respectively.

Note that the matrix elements in Tables~\ref{Smatrix}-\ref{Table l+2}
are of the form $\tme{n'l}{R}{nl}$, $\tme{n'l}{R}{n,l+1}$, or
$\tme{n'l}{R}{n,l+2}$, respectively, as this is the form that is most
naturally derived from applying the shift operators (see
Sec.~\ref{shift sec}).  However, one may express the $l$-changing matrix
elements in the form $\tme{n',l-1}{R}{nl}$ or $\tme{n',l-2}{R}{nl}$ by
making the appropriate substitution.  For instance, to obtain
$\tme{n',l-1}{r}{nl}$ from the matrix elements in Table~\ref{Table l+1},
one may simply make the substitution $l\rightarrow l-1$.  Alternatively, matrix
elements of the form $\tme{n',l-1}{R}{nl}$ or $\tme{n',l-2}{R}{nl}$
may be obtained by Hermitian conjugation.  For instance, to obtain
$\tme{n',l-1}{r}{nl}$ from the matrix elements in Table~\ref{Table l+1},
one may simply note that $\tme{n',l+1}{r}{nl}=\tme{nl}{r}{n',l+1}$,
since the matrix elements are real-valued, and make the interchange
$n\leftrightarrow n'$.  When applying Hermitian conjugation to
differential operators appearing in
Tables~\ref{Smatrix}-\ref{Table l+2}, recall that the operator $d/dr$
is anti-Hermitian (\textit{e.g.}, Ref.~\cite{sakurai1994:qm}).

\newcommand\T{\rule{0pt}{5ex}}
\newcommand\B{\rule[-2ex]{0pt}{0pt}}
\newcolumntype{M}{>{\T$}l<{$\B\hspace{.5cm}}}
\newcolumntype{T}{>{\T$}r<{$\B}}
\newcolumntype{L}{>{\hspace{.5cm}\T$}l<{$\B}}
\begin{center}

\begin{longtable}{MMT}
\caption{Matrix elements between functions in the $S_{nl}$ basis.\label{Smatrix}}\\
\hline\endfirsthead
\multicolumn{3}{c}{{\tablename\ \thetable{}
--continued from previous page}}\\

\hline
\endhead
\hline\multicolumn{3}{r}{{Continued on next page}}\\ 
\endfoot
\hline
\endlastfoot
\hline
\braket{n'l| r|nl}
&n'\ge n+2&0\\
&n'=n+1&-\frac{1}{2}((n+1)(n+2l+3))^{1/2}\\
&n'=n&n+l+\frac{3}{2}\\
&n'=n-1&-\frac{1}{2}(n(n+2l+2))^{1/2}\\
&n'\le n-2&0\\
\hline
\braket{n'l| r^2|nl}
&n'\ge n+3&0\\
&n'=n+2&\frac14((n+1)(n+2)(n+2l+3)(n+2l+4))^{1/2}\\
&n'=n+1&-(n+l+2)((n+1)(n+2l+3))^{1/2}\\
&n'=n&( n+l+\frac32)(n+l+2)+\frac12n(n+2l+2)\\
&n'=n-1&-(n+l+1)(n(n+2l+2))^{1/2}\\
&n'=n-2&\frac14(n(n-1)(n+2l+2)(n+2l+1))^{1/2}\\
&n'\le n-3&0\\
\hline
\braket{n'l|\frac{1}{ r}|nl}
&n'\ge n&\frac{1}{l+1}\left(\frac{n'!(n+2l+2)!}{n!(n'+2l+2)!}\right)^{1/2}\\
&n'\le n&\frac{1}{l+1}\left(\frac{n!(n'+2l+2)!}{n'!(n+2l+2)!}\right)^{1/2}\\
\hline
\braket{n'l|\frac{1}{ r^2}|nl}
&n'\ge n&{\left(\frac{2n'(2l+3)-2n(2l+1)+4l+6}{(l+1)(2l+1)(2l+3)}\right)}\left(\frac{n'!(n+2l+2)!}{n!(n'+2l+2)!}\right)^{1/2}\\
&n'\le n& {\left(\frac{2n(2l+3)-2n'(2l+1)+4l+6}{(l+1)(2l+1)(2l+3)}\right)}\left(\frac{n!(n'+2l+2)!}{n'!(n+2l+2)!}\right)^{1/2}\\
\hline
\braket{n'l|r\frac{d}{d r}|nl}
&n'\ge n+2&0\\
&n'=n+1&\frac12((n+1)(n+2l+3))^{1/2}\\
&n'=n&-\frac12\\
&n'=n-1&-\frac12(n(n+2l+2))^{1/2}\\
&n'\le n-2&0\\
\hline
\braket{n'l|\frac{d}{d r}|nl}
&n'\ge n+1&\left(\frac{n'!(n+2l+2)!}{n!(n'+2l+2)!}\right)^{1/2}\\
&n'=n&0\\
&n'\le n-1&- {\left(\frac{n!(n'+2l+2)!}{n'!(n+2l+2)!}\right)^{1/2}}\\
\hline
\braket{n'l|\frac{d^2}{d r^2}|nl}
&n'\ge n+1&-\frac{(2l+4)(2l+1)n-2l(2l+3)n'+(2l+3)(2l+2)}{(2l+3)(2l+1)}\left(\frac{n'!(n+2l+2)!}{n!(n'+2l+2)!}\right)^{1/2}\\
&n'=n&-\frac{4n(l+1)+2l+3}{(2l+3)(2l+1)}\\
&n'\le n-1&-\frac{(2l+4)(2l+1)n'-2l(2l+3)n+(2l+3)(2l+2)}{(2l+3)(2l+1)}\left(\frac{n!(n'+2l+2)!}{n'!(n+2l+2)!}\right)^{1/2}\\
\end{longtable}
\end{center}
\begin{center}
\begin{longtable}{MMT}
\caption{ \label{Table l+1}Matrix elements between functions in the $S_{n,l+1}$ and $S_{nl}$ bases.}\\
\hline\endfirsthead
\multicolumn{3}{c}{{\tablename\ \thetable{}
--continued from previous page}}\\

\hline
\endhead
\hline\multicolumn{3}{r}{{Continued on next page}}\\ 
\endfoot

\hline
\endlastfoot

\hline
\braket{n'l|n,l+1}
&n'\ge n+2&0\\
&n'=n+1&-\left(\frac{n+1}{n+2l+4}\right)^{1/2}\\
&n'\le n&(2l+3)\left(\frac{n!(n'+2l+2)!}{n'!(n+2l+4)!}\right)^{1/2}\\
\hline
\braket{n'l|r|n,l+1}
&n'\ge n+3&0\\
&n'=n+2&\frac12((n+1)(n+2))^{1/2}\\
&n'=n+1&-((n+1)(n+2l+4))^{1/2}\\
&n'=n&\frac12((n+2l+4)(n+2l+3))^{1/2}\\
&n'\le n-1&0\\
\hline
\braket{n'l|\frac1r|n,l+1}
&n'\ge n+1&0\\
&n'\le n&(2(n+1)-2n')\left(\frac{n!(n'+2l+2)!}{n'!(n+2l+4)!}\right)^{1/2}\\
\hline
\braket{n'l|\frac{d}{d r}|n,l+1}
&n'\ge n+2&0\\
&n'=n+1&\left(\frac{n+1}{n+2l+4}\right)^{1/2}\\
&n'\le n&(n'(2l+4)-n(2l+2)+1)\left(\frac{n!(n'+2l+2)!}{n'!(n+2l+4)!}\right)^{1/2}\\

\end{longtable}
\end{center}

\newcolumntype{M}{>{\T$}l<{$\B\hspace{.5cm}}}
\newcolumntype{T}{>{\T$}r<{$\B}}
\newcolumntype{L}{>{\hspace{.5cm}\T$}l<{$\B}}
\begin{center}
\begin{longtable}{MMT}
\caption{Matrix elements between functions in the $S_{n,l+2}$ and $S_{nl}$ bases.\label{Table l+2}}\\
\hline\endfirsthead
\multicolumn{3}{c}{{\tablename\ \thetable{}
--continued from previous page}}\\

\hline
\endhead
\hline\multicolumn{3}{r}{{Continued on next page}}\\ 
\endfoot
\hline
\endlastfoot
\hline
\braket{n'l|n,l+2}
&n'\ge n+3&0\\
&n'=n+2&\left(\frac{(n+1)(n+2)}{(n+2l+6)(n+2l+5)}\right)^{1/2}\\
^\mathrm{a}&n'\le n+1&c_{nn'}\left(\frac{n!(n'+2l+2)!}{n'!(n+2l+6)!}\right)^{1/2}\\
\hline
\braket{n'l|r|n,l+2}
&n'\ge n+4&0\\
&n'=n+3&-\frac12\left(\frac{(n+3)!}{n!(n+2l+6)}\right)^{1/2}\\
&n'=n+2&(n+3l+\frac{15}{2})\left(\frac{(n+2)!(n+2l+4)!}{n!(n+2l+6)!}\right)^{1/2}\\
&n'=n+1& {-\frac12(n(n+14)+6l(n+2l+9)+60)}\left(\frac{(n+1)(n+2l+3)!}{(n+2l+6)!}\right)^{1/2}\\
&n'\le n&{(l+2)(2l+3)(2l+5)}\left(\frac{n!(n'+2l+2)!}{n'!(n+2l+6)!}\right)^{1/2}\\
\hline 
\braket{n'l|r^2|n,l+2}
&n'\ge n+5&0\\
&n'=n+4&\frac14\left(\frac{(n+4)!}{n!}\right)^{1/2}\\
&n'=n+3&-\left(\frac{(n+2l+6)(n+3)!}{n!}\right)^{1/2}\\
&n'=n+2&\frac32\left(\frac{(n+2l+6)!(n+2)!}{n!(n+2l+4)!}\right)^{1/2}\\
&n'=n+1&-\left(\frac{(n+2l+6)!(n+1)}{(n+2l+3)!}\right)^{1/2}\\
&n'=n&\frac14\left(\frac{(n+2l+6)!}{(n+2l+2)!}\right)^{1/2}\\
&n'\le n-1&0\\
\hline 
\braket{n'l|\frac{1}{r^2}|n,l+2}
&n'\ge n+1&0\\
&n'\le n&\frac23(n-n'+1)(n-n'+2)(n-n'+3)\left(\frac{n!(n'+2l+2)!}{n'!(n+2l+6)!}\right)^{1/2}\\
\hline
\braket{n'l|\frac{d}{dr}|n,l+2}
&n'\ge n+3&0\\
&n'=n+2&-\left(\frac{(n+2)!(n+2l+4)!}{n!(n+2l+6)!}\right)^{1/2}\\
^\mathrm{b}&n'\le n+1&d_{nn'}\left(\frac{n!(n'+2l+2)!}{n'!(n+2l+6)!}\right)^{1/2}\\
\hline
\braket{n'l|\frac{d^2}{dr^2}|n,l+2}
&n'\ge n+3&0\\
&n'=n+2&\left(\frac{(n+2)!(n+2l+4)!}{n!(n+2l+6)!}\right)^{1/2}\\^\mathrm{c}
&n'\le n+2&f_{nn'}\left(\frac{n!(n'+2l+2)!}{n'!(n+2l+6)!}\right)^{1/2}
\end{longtable}
\begin{tablenotes}
\footnotesize{$^\mathrm{a}\,\,c_{nn'}=(2l+4)(2l+3)n-(2l+5)(2l+4)n'+(2l+4)(2l+3)$}\\
\footnotesize{$^\mathrm{b}\,\,d_{nn'}=n'[4n(l+2)^2+l(2l+9)+11-n'(2l+5)(l+3)]-(n+1)(2l+3)(n(l+1)-2)$}\\
\footnotesize{$^\mathrm{c}\,\,f_{nn'}=-\frac23\{n'[(l+3)n'((l+4)n'-3((l+2)n+1))+(l+2)n(3n(l+1)-6)-l(l+7)-9]$\\\hspace{2cm}$-n[(l+1)n(nl-9)-l(l+13)-9]+3l\}$}

\end{tablenotes}
\end{center}

\newpage
\providecommand{\APSLONG}{}
\providecommand{\ELSEVIER}{}


\end{document}